\documentclass[pdflatex,sn-mathphys-num,12pt]{sn-jnl}

\usepackage{graphicx}
\usepackage{multirow}
\usepackage{amsmath,amssymb,amsfonts}
\usepackage{amsthm}
\usepackage{mathrsfs}
\usepackage[title]{appendix}
\usepackage{xcolor}
\usepackage{textcomp}
\usepackage{manyfoot}
\usepackage{booktabs}
\usepackage{algorithm}
\usepackage{algorithmicx}
\usepackage{algpseudocode}
\usepackage{listings}
\usepackage{siunitx}
\usepackage[left]{lineno}

\usepackage{geometry}

\title{Predictive sufficiency as an information microscope for time-resolved coherent scattering}

\author*[1]{\fnm{Christian} \sur{Gutt}}
\email{christian.gutt@uni-siegen.de}

\author[1]{\fnm{Amir} \sur{Tosson}}
\author[3]{\fnm{Nimmi Das} \sur{Anthuparambil}}
\author[2]{\fnm{Jovita} \sur{Lukasik}}
\author[2]{\fnm{Michael} \sur{Möller}}

\affil*[1]{Department of Physics, University of Siegen,
57068 Siegen, Germany}

\affil[2]{Department of Electrical Engineering, University of Siegen,
57068 Siegen, Germany}

\affil[3]{Department of Physics, Indian Institute of Technology Madras,
Chennai 600036, Tamil Nadu, India}

\abstract{
X-ray photon correlation spectroscopy (XPCS) measures dynamics through the
evolution of coherent speckle patterns, most commonly summarized by intensity
correlation functions. These averages describe relaxation but can conceal how
speckle fluctuations evolve in time and across reciprocal space. Here we
introduce an information microscope that tests whether a chosen set of XPCS
observables retains the information needed to describe subsequent measured
scattering dynamics. A learned representation of the present speckle sequence
indicates when information remains outside the chosen description; an XPCS
Alphabet then searches for experimentally accessible observables that recover
it. In simulated Bragg scattering, speckle activity and anisotropy reveal
signatures of correlated deformation and intermittent slip. In a controlled
SAXS model, fluctuations of speckle activity provide sensitivity to the
spatial extent of cooperative particle motion. In a two-dimensional glass
former, retaining time-resolved speckle correlations and radial intensity
recovers information lost in the average relaxation function. The fluctuations
of these correlations become stronger during caging, without selecting a
single dominant correlated mode. This approach uses learning to guide the
physical analysis of coherent scattering, while keeping the resulting XPCS
observables measurable and interpretable.
}
\begin{document}

\maketitle

\begin{center}
    \small \today
\end{center}

\section*{Introduction}

A central problem in the analysis of time-resolved measurements is how to
represent a complex, evolving signal by a manageable set of observables. The
full measurement record is commonly reduced to correlation functions,
relaxation times, fluctuation amplitudes, order parameters, or related
descriptors. These quantities provide physical insight, but the reduction
can discard information relevant to subsequent measured dynamics. The
practical question is therefore whether the chosen observables retain the
information required to describe those dynamics.

This issue becomes particularly important for heterogeneous, intermittent,
aging, or otherwise complex dynamics. Two measurements can exhibit similar
mean relaxation while differing substantially in the temporal organization,
reciprocal-space distribution, or intermittency of their fluctuations. A
high-dimensional measurement may therefore contain dynamically relevant
information that is absent from a compact correlation function. Conversely,
a suitably chosen set of experimentally accessible observables may already
capture essentially all information relevant to a specified future
measurement.

X-ray photon correlation spectroscopy (XPCS) provides a particularly direct
setting in which to address this problem. Coherent X-ray scattering produces
speckle patterns whose temporal evolution encodes the dynamics of the
scattering structure
\cite{Sutton1991,GrubelMadsenRobert2008,PerakisGutt2020,Lehmkuehler2021}.
The recorded speckle movie contains the time-dependent scattering intensity,
whereas conventional XPCS extracts reduced observables such as the intensity
autocorrelation function $g_2(q,\tau)$, with scattering-vector magnitude $q$
and delay time $\tau$, two-time correlations, and measures of temporal
fluctuations and intermittent dynamics
\cite{Sutton2003,Duri2005,Trappe2007,Perakis2017,Wochner2009,Sanborn2011}.
These observables have provided detailed insight into relaxation and
collective motion across a broad range of materials and time scales. Yet it
is not generally tested whether a given reduction retains the information
relevant to subsequent measured dynamics.

This question is relevant across the range of current XPCS experiments.
Measurements of non-equilibrium and aging systems have revealed relaxation
beyond simple exponential behavior \cite{Madsen2010}. Atomic-scale XPCS
probes slow motion in amorphous materials
\cite{Ruta2013,Ruta2014,GiordanoRuta2016,Dallari2023,Shen2026}, while
resonant XPCS accesses fluctuations of ordered states
\cite{Chen2019,Klose2022}. Split-pulse, MHz, and pump--probe methods extend
coherent-scattering measurements towards faster and driven processes
\cite{Seaberg2017,Anthuparambil2026,Dargasz2026,Mangu2025}. Machine-learning
methods have also been used to denoise scattering data, classify dynamical
patterns, compare measurements with simulations, and extract dynamical
information \cite{Konstantinova2021,Konstantinova2022,Horwath2024,Cheng2026,
Timmermann2022,Andrejevic2024}. These developments make it increasingly
valuable to establish which information a conventional XPCS reduction
preserves and which it leaves behind.

Here we address this question through \emph{predictive sufficiency}. We use
experimentally accessible observables $X(t)$, evaluated from a present
measurement window, to describe a later scattering quantity $Y(t,\Delta)$.
The vector $X(t)$ may contain measurements at several scattering vectors and
correlation delays; it does not imply averaging over $q$. We then compare
this description with one that also includes a learned representation $Z(t)$
of the present speckle sequence. If $Z(t)$ improves the description of the
same future measurement, the chosen XPCS observables have left accessible
information unused. This practical test is related to predictive-state and
information-based approaches \cite{ShaliziCrutchfield2001,
LittmanSuttonSingh2001,TishbyPereiraBialek1999,BialekNemenmanTishby2001,
Creutzig2009,WangTiwary2021,Schmitt2024}. It does not require the speckle
movie to determine a unique microscopic state or predict individual
stochastic events.

We call this approach an \emph{information microscope} because the learned
representation is a diagnostic, not the intended final physical observable.
Once a missing contribution is detected, we search an \emph{XPCS Alphabet} of
quantities constructed from scattering intensities: correlations resolved in
wavevector and delay, short-time intensity changes, spatial variations of
speckle activity, and measures of diffraction-pattern shape. Candidates are
chosen for their ability to improve the description of subsequent XPCS
measurements. The observable search uses scattering intensities; controlled
simulations then allow us to examine the physical sensitivity of the
selected quantities. The benchmark-specific encoder training and model
selection are described separately in Methods and Supplementary Information.

We illustrate the approach in three scattering settings. Around a Bragg
reflection, we examine how changes in speckle activity and anisotropy encode
correlated lattice deformation and intermittent slip
\cite{RobinsonHarder2009,DAnna2026,Carr2021,Sheyfer2024,Carnis2019,
Clark2015,Orr2023}. In a controlled SAXS model, we distinguish the amplitude
of collective displacements from the spatial extent of cooperative motion.
Finally, in a dense two-dimensional glass, we identify which time-resolved
speckle correlations and intensity measures improve the description beyond
the mean relaxation, and examine their fluctuations as caging develops. The
three examples show how information retained in a coherent-scattering movie
can guide the choice of physically interpretable XPCS observables.

\section*{Results}

\subsection*{A predictive description of XPCS dynamics}

We describe the present scattering measurement by a set of experimentally
accessible XPCS observables \(X(t)\) and define a future target
\(Y(t,\Delta)\) from scattering data acquired at later times. Here \(t\)
marks the end of the present observation window and \(\Delta\) specifies
the separation to the later measured target, according to each benchmark. Depending
on the system, the target may describe a later intensity correlation or
another measured scattering quantity; its definition and time window are
specified for each benchmark. A regression model \(\hat Y=f(X)\) is trained
on one set of speckle sequences and tested on independent data. We quantify
prediction quality by the coefficient of determination \(R^2\).

To assess whether the chosen observables capture the information relevant
to the future measurement, we compare predictions from \(X(t)\) alone with
those obtained after adding a learned representation \(Z(t)\) of the same
present speckle movie. The gain
\(\Delta R^2=R^2([X,Z]\rightarrow Y)-R^2(X\rightarrow Y)\)
measures the additional predictive contribution of the movie representation
under the tested models. A statistically resolved positive gain motivates
a search for further measurable XPCS observables that can account for it.

\begin{figure*}[t]
    \centering
    \includegraphics[width=\textwidth]{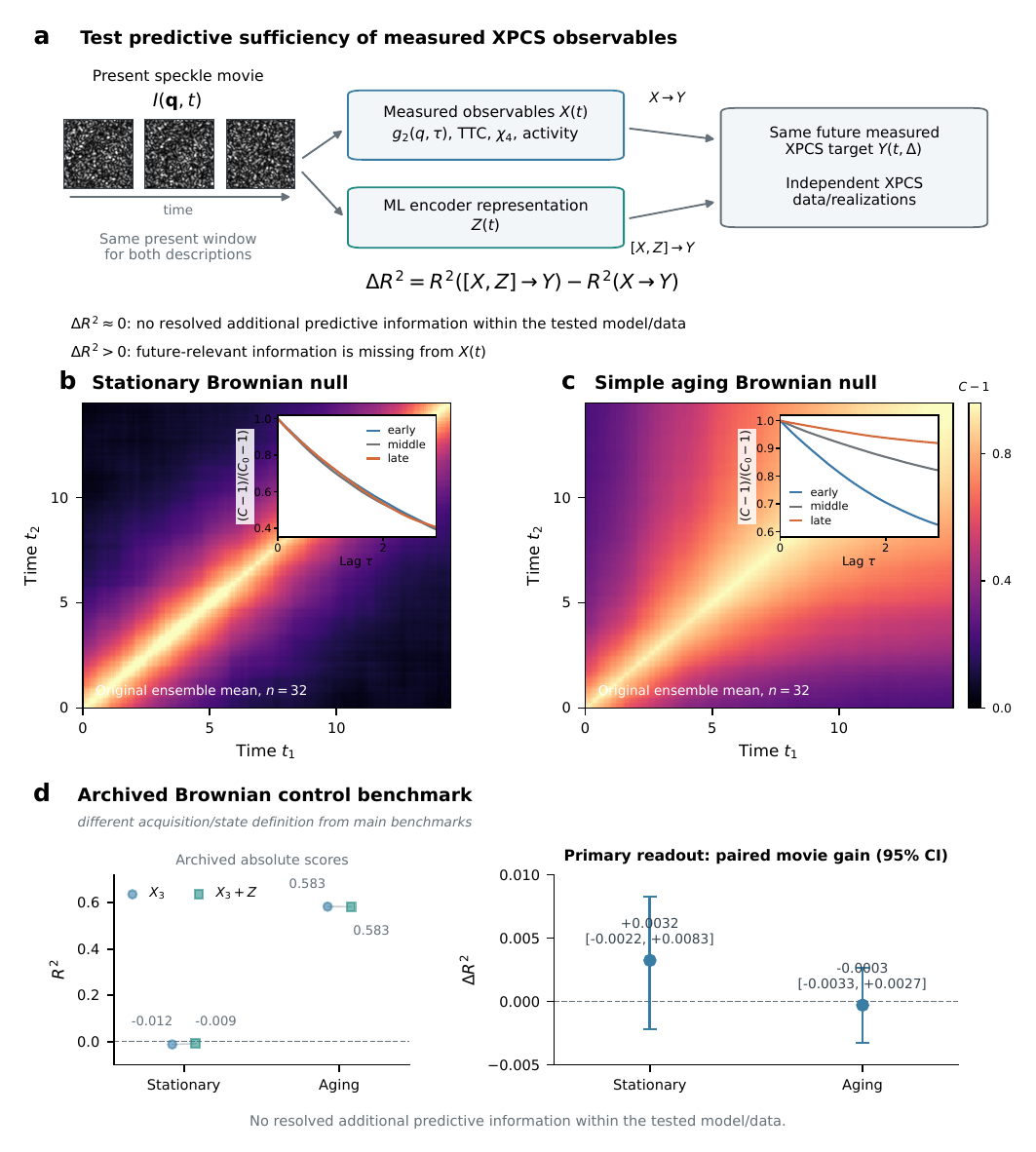}
    \caption{\textbf{Predictive description of XPCS dynamics and Brownian controls.}
    \textbf{a,} Present speckle patterns are characterized by XPCS observables \(X(t)\) and compared with subsequent scattering dynamics \(Y(t,\Delta)\); \(Z(t)\) represents additional information from the same speckle sequence.
    \textbf{b,c,} Independently generated, illustrative two-time correlations for stationary and aging Brownian motion; these panels do not supply the quantitative prediction results in panel d.
    \textbf{d,} Ridge prediction of subsequent correlations using \(X_3\), comprising mean intensity correlations, fluctuations of speckle decorrelation, and their time-resolved averages. The additional speckle information gives no resolved improvement in either case. Error bars show archived 95\% paired whole-movie confidence intervals.}
    \label{fig:predictive_sufficiency}
\end{figure*}

Stationary and smoothly aging Brownian dynamics provide simple controls
(Fig.~\ref{fig:predictive_sufficiency}). Panel a compares predictions
from the measured observables \(X\) alone with those including the learned
speckle representation \(Z\), using the same later measured target.
The two-time correlations in panels b and c illustrate stationary relaxation
and gradual aging, respectively. In the stationary system, the ensemble-mean
relaxation is reproducible, but fluctuations in a later finite measurement
window are only weakly predictable from the preceding window at the tested
horizon. In the aging system, the gradual slowing is already reflected in
the measured intensity correlations and provides predictability. When the explicit description includes
correlations at several wave vectors and delay times, their temporal
fluctuations, and correlations evaluated over successive intervals of the
present measurement, adding the learned speckle representation yields no
statistically resolved improvement in either case (Fig.~\ref{fig:predictive_sufficiency}d).
These controls illustrate two distinct outcomes of the same test: little
predictable information in the measured future, or predictable dynamics
already captured by conventional XPCS quantities. They do not establish
information-theoretic sufficiency beyond the tested models and data.

\subsection*{Bragg speckles reveal information beyond a compact XPCS description}

\begin{figure*}[t]
    \centering
    \includegraphics[width=\textwidth]{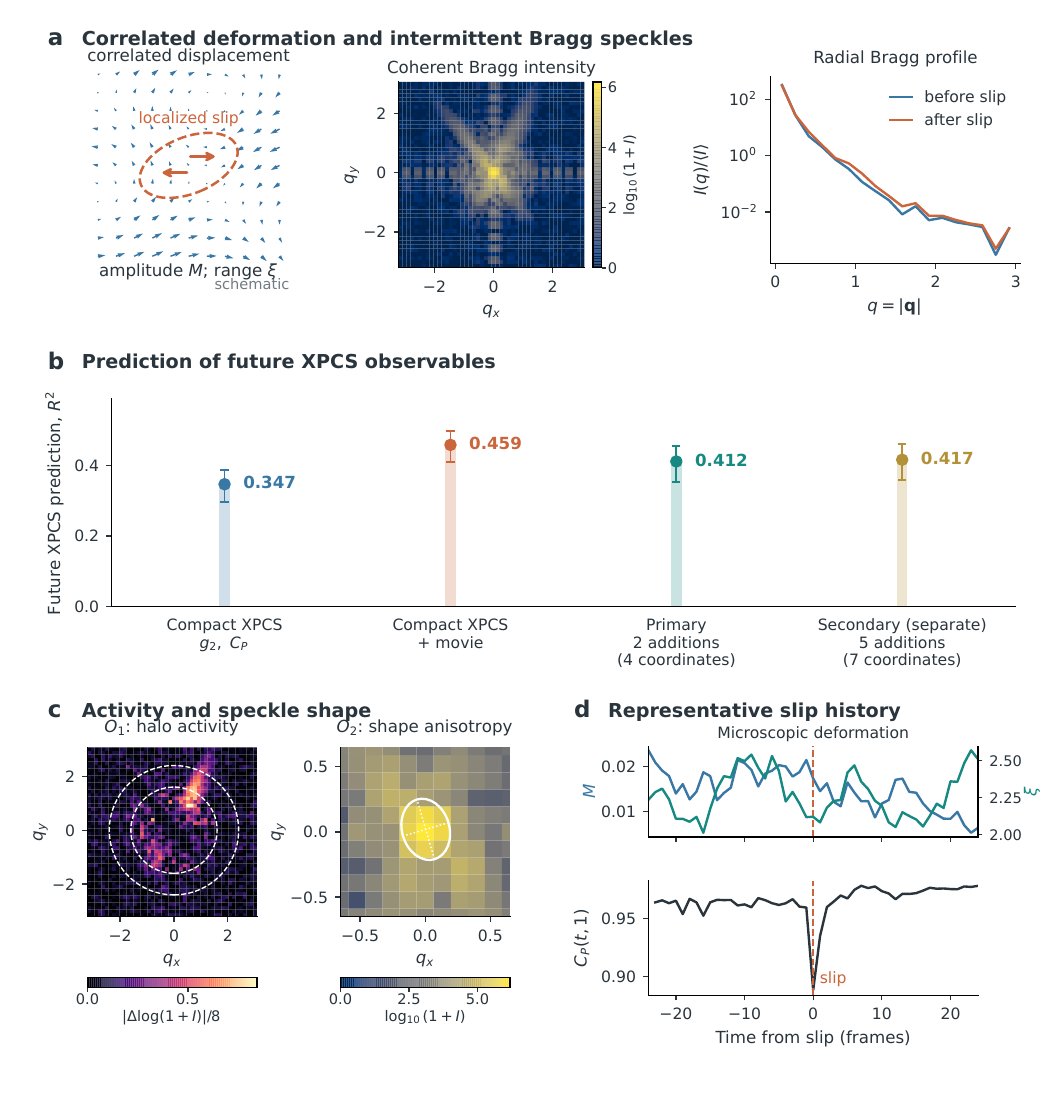}
    \caption{\textbf{Bragg speckles reveal signatures of correlated deformation and intermittent slip.}
    \textbf{a,} Schematic correlated lattice displacements and localized slip, with the resulting coherent Bragg intensity and radial intensity profile before and after a slip.
    \textbf{b,} Prediction of later Bragg-XPCS correlations using the compact description \(g_2\) and \(C_{\rm P}\), the additional learned speckle representation, and two or five observables identified by the XPCS Alphabet. The five-addition state is selected separately from the two-addition state. Error bars indicate 95\% whole-movie bootstrap intervals.
    \textbf{c,} Short-time intensity changes in the outer Bragg halo and anisotropy of the intensity distribution, illustrating the first two Alphabet observables.
    \textbf{d,} Representative evolution of simulated deformation amplitude \(M\), spatial correlation length \(\xi\), and short-delay speckle correlation \(C_{\rm P}\) around a slip event.}
    \label{fig:bragg}
\end{figure*}

We first consider coherent scattering around a Bragg reflection, where even small
lattice displacements modify the phase of the scattered wave and redistribute
intensity within the speckle pattern
(Fig.~\ref{fig:bragg}a)
\cite{RobinsonHarder2009,DAnna2026,Carr2021,Sheyfer2024,
Carnis2019,Clark2015,Orr2023}.
For a reciprocal-lattice vector \(\mathbf G\) and a small deviation
\(\mathbf q\) from the Bragg position, the simulated coherent amplitude is
\[
A(\mathbf G+\mathbf q,t)
\propto
\int d\mathbf r\,w(\mathbf r)\,
\exp\!\left[i\mathbf G\cdot\mathbf u(\mathbf r,t)\right]
\exp(-i\mathbf q\cdot\mathbf r),
\qquad I(\mathbf G+\mathbf q,t)=|A(\mathbf G+\mathbf q,t)|^2.
\]
Here \(w(\mathbf r)\) denotes the spatial illumination envelope.
The displacement field consists of
reversible, spatially correlated distortions and permanent changes caused by
localized slip events:
\[
\mathbf u(\mathbf r,t)
=\mathbf u_{\mathrm{el}}(\mathbf r,t)
+\mathbf u_{\mathrm{pl}}(\mathbf r,t).
\]
The elastic displacement field evolves according to
\[
\mathbf u_{\mathrm{el}}(\mathbf r,t+1)
=0.82\,\mathbf u_{\mathrm{el}}(\mathbf r,t)
+[M(t)+B(t)]\,\mathbf f_{\xi(t)}(\mathbf r,t),
\]
where \(\mathbf f_{\xi}\) is a normalized, spatially correlated Gaussian
random vector field with correlation length \(\xi\). The amplitude \(M\)
sets the strength of the reversible displacements, while \(B\) provides
a transient enhancement following a slip event. The factor 0.82 describes
the persistence of earlier deformations. Localized slip events additionally
produce permanent increments in the plastic displacement field. The model
thus generates continuous speckle fluctuations from correlated lattice
motion and abrupt changes from intermittent rearrangements, as encountered
in Bragg-XPCS. Predictions use only the simulated detector intensities;
the displacement and event variables are used subsequently for physical
interpretation.

We initially characterize the measured Bragg dynamics by two correlations
at a delay of four frames: the usual intensity autocorrelation \(g_2\) and
a spatial Pearson correlation \(C_{\mathrm P}\) between speckle patterns.
While \(g_2\) measures intensity decorrelation, \(C_{\mathrm P}\) emphasizes
changes in the spatial speckle configuration after removing the mean
intensity and contrast of each frame. We ask how well these two quantities,
measured over a short present window, predict their counterparts in a later,
non-overlapping window separated by 16 frames.

The two correlations predict the future Bragg-XPCS observables with
\(R^2=0.347\). Including information learned directly from the present
speckle sequence increases the prediction score to \(R^2=0.459\)
(Fig.~\ref{fig:bragg}b). Thus, the evolving speckle pattern contains
information about subsequent dynamics that is not captured by conventional
intensity decorrelation and spatial speckle similarity alone. This gain
reflects information missing from the chosen experimental description,
not an ability to predict every future rearrangement: stochastic deformation
and slip events also limit the reproducibility of future speckle fluctuations
(see Supplementary Information). 

To recover this additional information in experimentally accessible terms,
we use the XPCS Alphabet: a library of candidate observables built from
speckle-intensity changes, temporal fluctuations, reciprocal-space variations,
and pattern-shape measures. Starting from the two conventional correlations,
we select additions according to how much they improve prediction of the
same future XPCS observables. Importantly, this search uses only measured
scattering intensities, not the known microscopic deformation or slip variables.
It therefore asks which physical features of the speckle movie are missing
from the compact correlation description.

The search first identifies two complementary quantities
(Fig.~\ref{fig:bragg}c). The first, \(O_1\), characterizes strong, short-time intensity
changes in the outer Bragg halo relative to activity in other
reciprocal-space regions. The second, \(O_2\), captures anisotropy of the
intensity distribution during the observation window. Together, they retain
information about where intensity changes occur and how the overall Bragg
pattern changes shape---features not distinguished by the two initial
correlation measures.

Adding these observables increases the prediction score from \(R^2=0.347\)
to \(0.412\), recovering much of the improvement obtained from the learned
movie representation (\(R^2=0.459\)). Three further short-delay and
\(q\)-resolved observables bring only a small additional improvement, to
\(R^2=0.417\). The learned representation still improves prediction beyond
this enriched XPCS description, indicating that some information in the
speckle sequence remains unaccounted for.

Having identified scattering observables that improve the prediction of
future Bragg-XPCS correlations, we next ask what these observables reveal
about the underlying motion. The simulated deformation field and slip history
are known, allowing a separate test of how well microscopic properties can be
inferred from the measured XPCS description. These microscopic variables were
not used in selecting the Alphabet observables. Figure~\ref{fig:bragg}d
illustrates a representative slip event: the short-delay speckle correlation
changes sharply near the event, while the deformation amplitude and correlation
length evolve over the same interval. The figure illustrates the dynamics to be
inferred, not the accuracy of the inference.

The enriched XPCS description is particularly sensitive to intermittent
motion: the reconstruction score for the number of recent slip events
increases from $R^2=0.366$ to $0.517$, and that for the time since the
last event from $0.229$ to $0.317$. For the separately selected
\emph{five-observable} description, the score for the reversible-deformation
amplitude $M$ increases from $0.096$ to $0.274$, while the spatial correlation
length $\xi$, weakly represented by the initial correlations, reaches
$R^2=0.270$. These results do not establish a unique reconstruction of
individual slip events or displacement fields. Rather, the spatial distribution
of speckle activity and changes in the Bragg-pattern shape carry complementary
signatures of correlated deformation and recent slip history.

This conclusion is not specific to a linear prediction model. Nonlinear
prediction and conventional covariance-based descriptions also recover part
of the missing information, but neither removes the additional contribution
of the learned speckle representation. Details of these comparisons and their
statistical uncertainties are given in the Supplementary Information.

\subsection*{A controlled SAXS model separates dynamical activity from cooperativity}

\begin{figure*}[t]
    \centering
    \includegraphics[width=\textwidth]{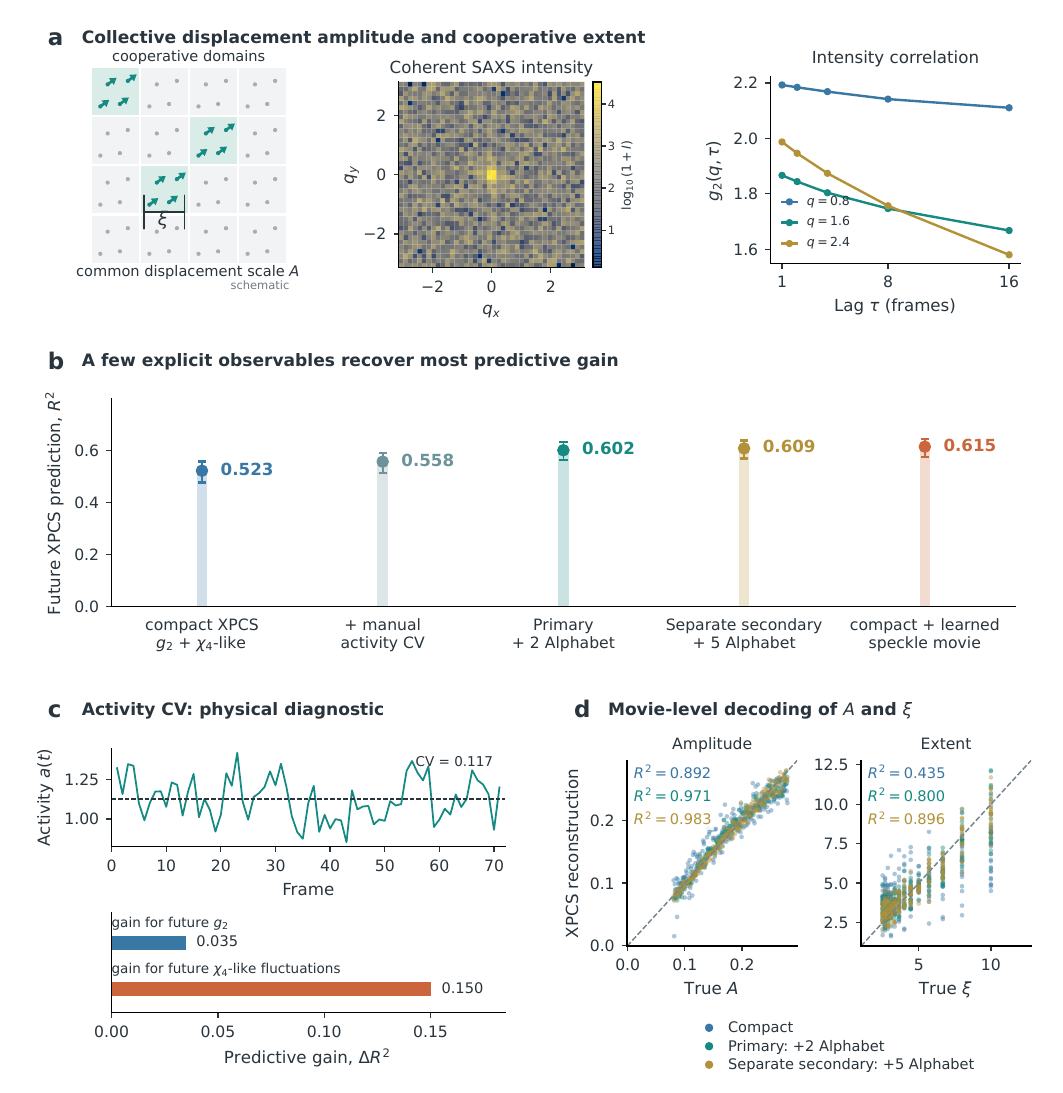}
    \caption{\textbf{XPCS fluctuations reveal the amplitude and extent of collective motion.}
    \textbf{a,} Model of intermittent collective displacements within domains of size
    \(\xi\) and amplitude \(A\), together with a representative coherent SAXS
    pattern and intensity correlations \(g_2(q,\tau)\).
    \textbf{b,} Prediction of subsequent XPCS correlations and fluctuation amplitudes
    from the initial \(g_2\) and \(\chi_4\)-like observables, an additional
    manually defined speckle-activity measure, and two or five observables
    selected by the XPCS Alphabet. The full learned speckle representation
    provides the comparison.
    \textbf{c,} Time-dependent speckle activity and its coefficient of variation
    (top), and the additional predictive contribution for future mean
    relaxation and \(\chi_4\)-like fluctuations (bottom).
    \textbf{d,} Reconstruction of the collective displacement amplitude \(A\)
    and cooperative domain size \(\xi\) from the compact and enriched XPCS
    descriptions. Dashed lines indicate exact agreement. Error bars in the
    prediction and reconstruction panels represent the reported uncertainty
    estimates (see Methods).}
    \label{fig:axi}
\end{figure*}

We next consider a model of coherent SAXS from particles that undergo
independent Brownian motion together with intermittent collective displacements
(Fig.~\ref{fig:axi}a). The collective motion acts within square spatial
domains of linear size \(\xi\): particles in the same active domain receive
the same displacement, whereas particles in different domains move
independently. Its amplitude \(A\) is the standard deviation of each
Cartesian component of this common displacement. Varying \(A\) and \(\xi\)
independently therefore allows us to distinguish how far particles move
from how many particles move together. In an XPCS experiment these effects
can both alter speckle decorrelation, making their separate identification
from conventional correlations nontrivial.

As a compact description of the measured dynamics, we use the intensity
autocorrelation \(g_2(t,\tau=16)\) and a \(\chi_4\)-like measure of
the fluctuations of the time-resolved correlation, both averaged over the
measured \(q\) range. We ask how well their values in a 72-frame observation
window predict the same quantities in the following non-overlapping window.
The mean correlation characterizes the overall relaxation, while its
fluctuations reflect how strongly the dynamics vary during the measurement.

These conventional XPCS quantities predict the future correlations and
fluctuation amplitudes with \(R^2=0.523\). Adding a representation learned
from the full present speckle sequence improves the prediction to
\(R^2=0.615\) (Fig.~\ref{fig:axi}b). The improvement is markedly larger
for the future \(\chi_4\)-like quantity (\(\Delta R^2=0.150\)) than for
the mean correlation (\(\Delta R^2=0.035\)). Thus, much of the average
relaxation is already captured by the compact XPCS description, whereas
the speckle movie contains additional information about the future strength
of dynamical fluctuations.

This points to the time dependence of speckle activity as a useful place
to look for the missing information. We first make a simple, physically
motivated test: from the changes in detector intensity between successive
frames, we calculate an activity signal and measure how strongly it varies
over the present observation window. Adding its coefficient of variation
(defined in Methods) raises the prediction score to \(R^2=0.558\)
(Fig.~\ref{fig:axi}b); the activity trace is shown in panel c.
The remaining predictive gain from adding \(Z\) to this enriched
measured state decreases from \(0.093\) to \(0.058\). This shows that
the intermittency of measured intensity changes
already accounts for an appreciable part of the information missed by the
mean correlation and its fluctuation amplitude.

We then broaden this targeted test into a systematic search using the
XPCS Alphabet. Rather than prescribing one activity measure, the Alphabet
generates candidate observables from experimentally available speckle
intensities, their short-time changes, and their dependence on scattering
vector and time. Each candidate is tested for whether it improves prediction
of the same future XPCS quantities beyond the two initial correlations.
The observable search uses the measured scattering intensities, not
\(A\), \(\xi\), or particle trajectories; the separate historical choice
of encoder architecture did use these simulated parameters (see SI). In this way, the learned movie representation
first identifies an information deficit, and the Alphabet searches for
measurable scattering quantities that can account for it.

The first selected observable combines temporal variations in the overall
strength of frame-to-frame intensity changes with a fluctuation measure
from a fixed \(q\) region. It therefore retains both \emph{when} strong
speckle activity occurs and \emph{which} part of reciprocal space
contributes. The second characterizes how the spread of intensity changes
across detector pixels varies over time. In contrast to a single averaged
correlation, these quantities distinguish speckle sequences with similar
mean decorrelation but different histories of fluctuating activity.
They do not, by themselves, measure the real-space size of a cooperative
domain.

Adding these two automatically selected observables increases the
prediction to \(R^2=0.602\), leaving an additional movie contribution of
only \(0.014\). An independently selected, larger set of five additions, which also
retains spatial dispersion of intensity changes between detector patches
and additional \(q\)-resolved intensity-change summaries, gives
\(R^2=0.609\) and a remaining contribution of \(0.008\). Thus,
most of the predictive improvement initially found in the full speckle
movie can be recovered through a small set of explicit quantities
calculated from the measured scattering intensities. The detailed
definitions and selection tests are provided in the Supplementary
Information.

The model now allows us to examine what physical information these
observables have recovered. Only after the scattering-based selection
is complete do we use the known values of \(A\) and \(\xi\) to test how
well they can be inferred from the measured XPCS descriptions
(Fig.~\ref{fig:axi}d). The collective displacement amplitude \(A\)
is already well constrained by the compact correlations
(\(R_A^2=0.892\)); the enriched descriptions raise this to
\(0.971\) and \(0.983\). The contrast is much stronger for the
cooperative domain size: the compact description gives only
\(R_\xi^2=0.435\), whereas the two- and five-observable enrichments
reach \(0.800\) and \(0.896\), respectively.

The central result is therefore not simply a better prediction of
future correlations. Although the original observables already capture
much of the displacement amplitude, the additional time- and
\(q\)-resolved speckle fluctuations provide sensitivity to the spatial
extent of collective motion. This sensitivity is established through
the controlled model comparison; it should not be confused with a
direct measurement of \(\xi\) from an individual scattering observable.
The XPCS Alphabet thus turns information detected in the full speckle
movie into interpretable experimental signatures of cooperative dynamics.

\subsection*{A dense two-dimensional glass exposes information beyond the mean XPCS relaxation}

\begin{figure*}[t]
\centering
\includegraphics[width=12cm]{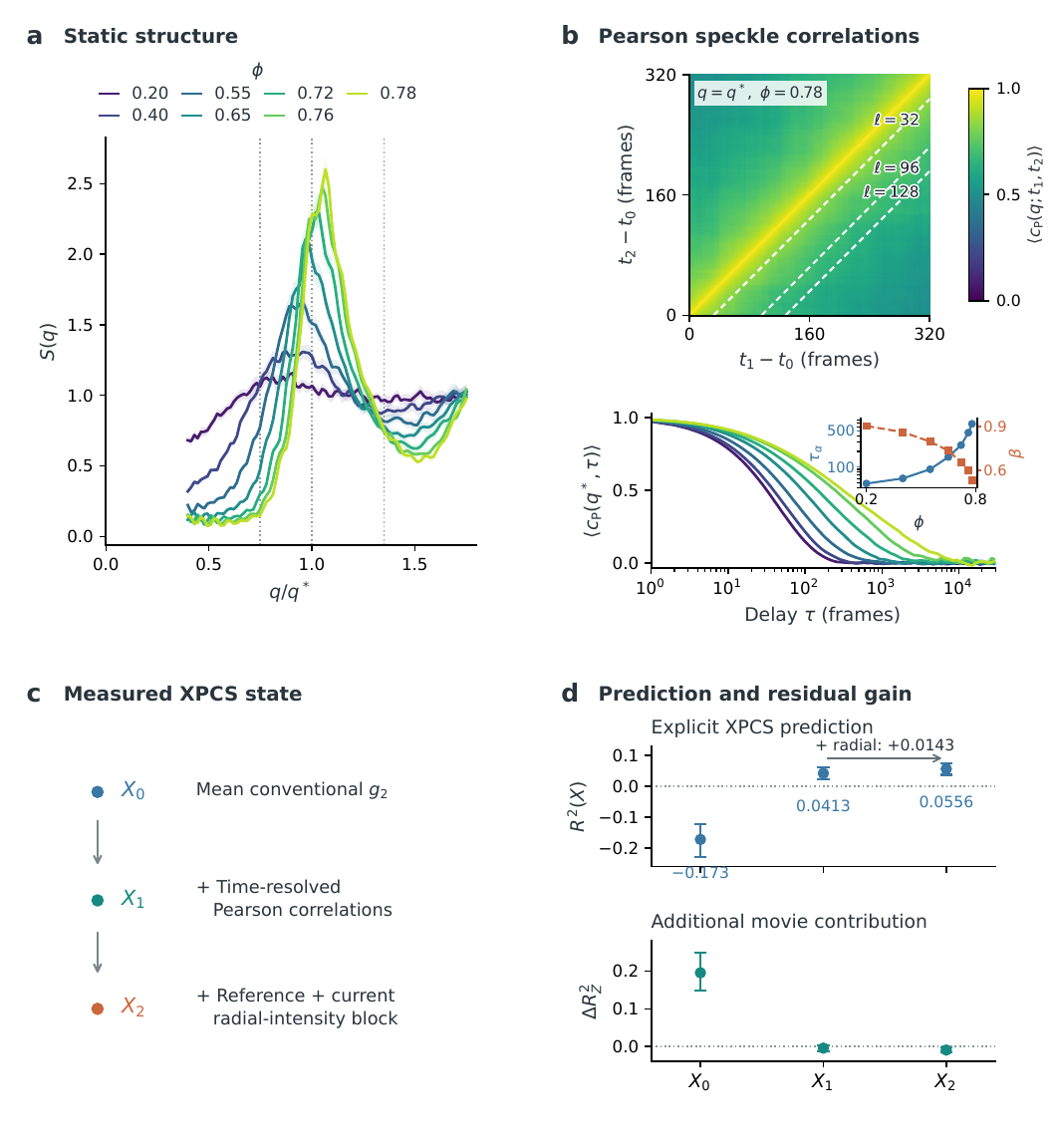}
\caption{\textbf{From conventional glass dynamics to an explicit XPCS-state description.}
\textbf{a,} Static structure factor \(S(q)\) across packing fractions. The
 dotted lines indicate the representative wavevectors used in the predictive
 XPCS analysis; \(q^*=2\pi\) is the fixed reciprocal-space reference of the simulation.
\textbf{b,} Representative two-time Pearson speckle-pattern correlation at
 \(q=q^*\) for the densest state (top); independently averaged Pearson
 relaxation curves across packing fractions (bottom). The inset shows the increasing relaxation time
 \(\tau_\alpha\) and decreasing KWW exponent \(\beta\).
\textbf{c,} Cumulative measured XPCS-state ladder, starting from the mean
 \(g_2\) (\(X_0\)), followed by time-resolved speckle-pattern correlations
 (\(X_1\)) and reference-plus-current radial-intensity information (\(X_2\)).
\textbf{d,} Predictive information along the same ladder. Top: prediction
 score \(R^2(X)\) of the explicit XPCS state; adding the reference-plus-current
 radial-intensity block further improves prediction from \(X_1\) to \(X_2\).
 Bottom: additional predictive contribution \(\Delta R_Z^2\) of the full
 present speckle history beyond each explicit state. The large additional
 contribution beyond \(g_2\) disappears once the time-resolved speckle-pattern
 correlations are retained.}
\label{fig:glass_conventional}
\end{figure*}

Our final example is a two-dimensional 50:50 mixture of soft disks with a
diameter ratio of \(1.4\). Upon increasing the packing fraction from
\(\phi=0.20\) to \(0.78\), the static scattering remains liquid-like,
while the intensity correlations show the characteristic slowing associated
with caging. The relaxation time \(\tau_\alpha\) increases and the KWW
exponent \(\beta\) decreases, indicating a broader relaxation. Particle
mean-squared displacements independently exhibit an intermediate-time cage
regime (see Supplementary Information). The corresponding structure factors and XPCS relaxation curves are shown
in Fig.~\ref{fig:glass_conventional}a,b; the predictive-state comparison
appears separately in panels c,d.

\begin{figure*}[p]
\centering
\includegraphics[width=0.68\textwidth]{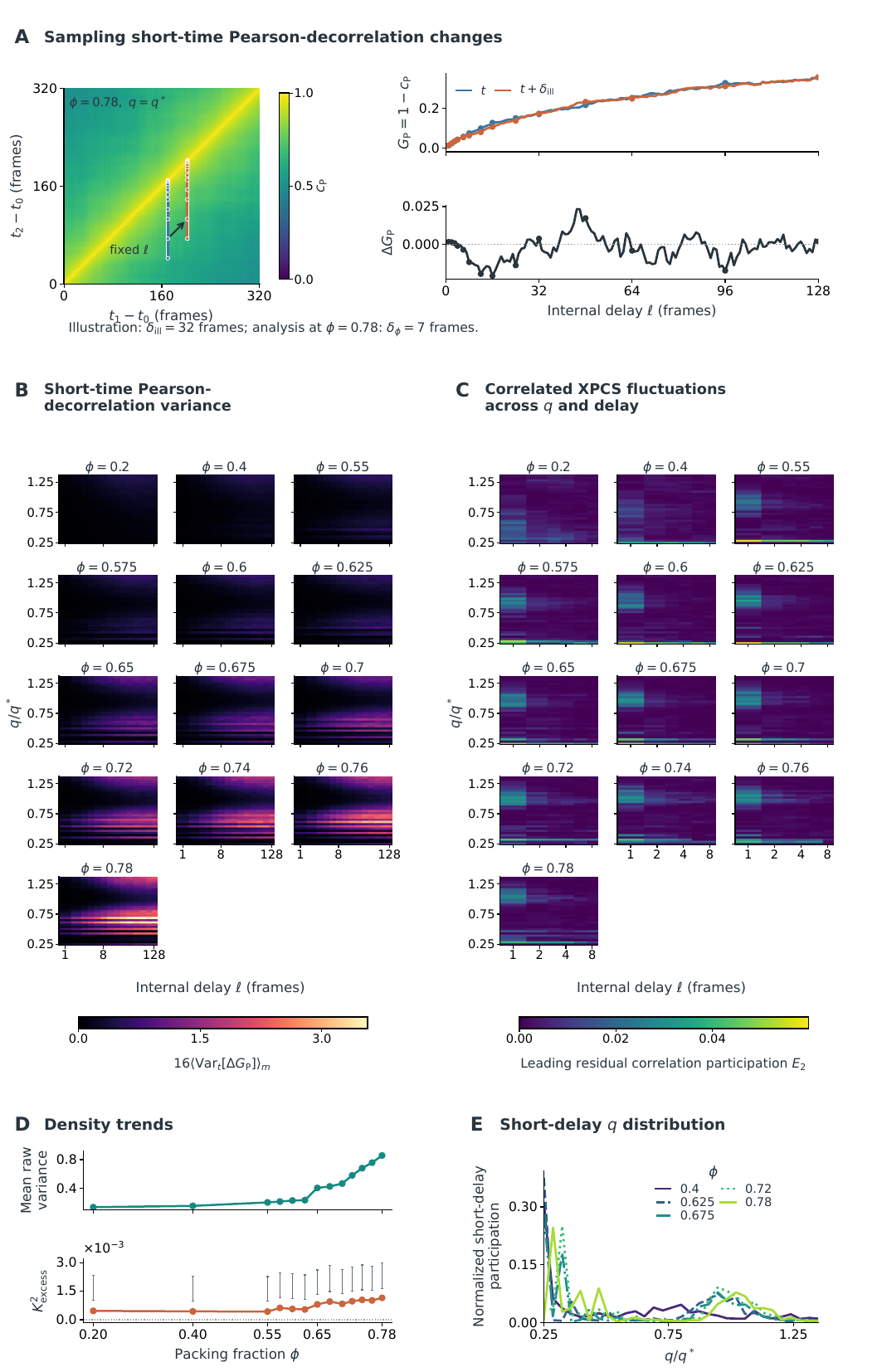}
\caption{\textbf{Fluctuations and correlations within the time-resolved XPCS state.}
\textbf{a,} Short-time changes in the Pearson speckle-pattern decorrelation
\(G_{\rm P}(q,t,\ell)\), obtained by comparing two cuts through the
TTC at time origins separated by \(\delta_\phi\), while keeping the internal
delay \(\ell\) fixed. The illustration uses \(\delta=32\) frames;
the analysis at \(\phi=0.78\) uses \(\delta_\phi=7\) frames.
\textbf{b,} Fluctuation amplitude of these short-time decorrelation changes
across scattering vector and internal delay, shown on a common color scale.
The \(q\)-dependence resembles the relaxation-rate modulation
\(q^2/S(q)\).
\textbf{c,} Distribution of the leading correlated fluctuations across
\(q\) and short delays, after removing specified mean-relaxation contributions
and normalizing channel variances. Colors indicate participation in the
leading two correlation directions, not fluctuation amplitude.
\textbf{d,} Evolution with packing fraction of the mean raw fluctuation variance (top) and the null-corrected squared residual cross-channel correlation strength $K_{\mathrm{excess}}^2=K_{\mathrm{rms,obs}}^2-\langle K_{\mathrm{rms,null}}^2\rangle_{\mathrm{null}}$ (bottom).
\textbf{e,} Reciprocal-space profile of the leading correlated sector,
summed over six short internal delays. Its weight remains broadly distributed
in \(q\), without a sharply selected wavevector.}
\label{fig:glass_covariance_summary}
\end{figure*}

We first characterize the dynamics by the time-averaged intensity correlation
\(g_2(q,\tau)\) at \(q/q^*=0.75,1.0,1.35\), where the fixed reference
\(q^*=2\pi\) sets the reciprocal-space scale. This compact description
captures the average relaxation but removes variations between individual
time origins. In the prediction test of Fig.~\ref{fig:glass_conventional},
\(X_0\) is a fixed-prefix mean-\(g_2\) calibration vector, repeated within
each movie, rather than a moving-window correlation estimate. Adding the
learned representation of the present speckle history increases the
prediction score by \(\Delta R_Z^2=0.196\). This demonstrates that the
fixed mean-correlation reference omits information relevant to subsequent
scattering dynamics; by itself, it does not establish that every
conventional time-resolved \(g_2\) description is insufficient.

The XPCS Alphabet helps identify which additional parts of the measured
speckle signal are useful. In a separate analysis using only earlier
scattering frames to establish the mean-\(g_2\) reference, it first selects
the angular spread of the instantaneous speckle intensities at three
scattering vectors. It subsequently selects combinations of speckle-pattern
correlations at short and longer delays. These quantities describe two
familiar properties of coherent scattering: intensity variations around a
ring and the loss of speckle similarity with time. Their calculation
requires neither particle coordinates nor a supplied packing-fraction label.

These five scattering quantities improve prediction of the later XPCS
correlations in all three independent test cohorts. For the largest cohort,
\(R^2\) increases from \(-0.182\) for the mean-correlation description to
\(0.041\) with the selected observables. The improvement persists when the
future correlation is formed entirely from frames recorded after the present
observation window. The full movie still provides a small additional
improvement for the primary target. Across repeated searches, the precise
selected quantities vary, but intensity-amplitude and time-resolved
correlation measures recur (details in the Supplementary Information).

The selected correlations motivate retaining the time dependence of the
speckle pattern rather than reducing it to a mean relaxation curve. We
therefore examine the ring-wise Pearson correlation
\begin{equation}
c_{\rm P}(q,t,\ell)=
\mathrm{corr}_{\theta}\left[I(q,\theta,t),I(q,\theta,t-\ell)\right],
\end{equation}
and its corresponding decorrelation
\(G_{\rm P}(q,t,\ell)=1-c_{\rm P}(q,t,\ell)\).
The following comparison is the separate, physically defined XPCS-state
analysis shown in Fig.~\ref{fig:glass_conventional}; its prediction scores
are not combined with those of the Alphabet search.

Pearson normalization removes the instantaneous mean intensity and overall
intensity scale of each ring, emphasizing changes in the registered speckle
configuration. The angular coordinate labels detector positions along a
constant-\(q\) ring; it is not used here to determine orientational order.
We calculate pairwise Pearson correlations among 15 speckle patterns sampled
from the preceding 128 frames. Together with the fixed mean-\(g_2\)
reference, these values form \(X_1=[g_2,G_{\rm P}]\), which retains the
temporal organization of the recorded speckle correlations.

In the original Figure~\ref{fig:glass_conventional} test, \(X_1\) gives
\(R^2=0.041\), while adding the learned speckle representation gives
\(R^2=0.037\), with no positive incremental gain. This indicates that the
explicit time-resolved Pearson description recovers much of the predictive
contribution missing from the fixed mean-correlation reference, although
the absolute predictability of later correlations remains low.

To distinguish temporal resolution from Pearson normalization, we also
examined moving-window conventional intensity correlations and matched
pairwise conventional two-time correlations (Supplementary Information).
Moving-window conventional correlations improve on the fixed-prefix
reference, but the learned representation retains a positive incremental
contribution in the tested summaries. At matched correlation endpoints
and state dimension, Pearson correlations provide a resolved predictive
advantage over conventional pairwise correlations when the instantaneous radial scattering intensity is excluded.

We next retain the instantaneous radial scattering intensity
\(I_{\mathrm{rad}}(q,t)=\langle I(q,\theta,t)\rangle_\theta\),
whose fluctuating amplitude is removed by Pearson normalization. In the original Fig.~\ref{fig:glass_conventional} analysis,
\(X_2\) adds both a density-dependent reference scattering profile and
the instantaneous radial intensity relative to that reference, improving prediction from
\(R^2=0.041\) to \(0.056\). This comparison alone cannot separate the
two contributions. Matched controls without a supplied density-dependent
reference show that the instantaneous radial intensity itself improves prediction,
without explicitly supplying a density-dependent mean profile.
Once the same instantaneous radial-intensity information is supplied
to both pairwise correlation descriptions, the Pearson advantage is no longer resolved.
The remaining contribution of the learned representation to conventional
pairwise correlations plus instantaneous radial intensity is also unresolved, although
predictive equivalence is not established (Supplementary Information).

These comparisons show that predictive information absent from a fixed
mean-relaxation reference can be recovered by explicit time-resolved
correlations together with instantaneous radial scattering intensity. Pearson
correlations are advantageous in a matched comparison without this intensity information, but their advantage is not resolved
once the same instantaneous radial-intensity measurements are available
to both representations. The results therefore
do not identify a uniquely predictive Pearson-specific or hidden glass
coordinate. Instead, they motivate examining how the fluctuations of the
measured time-resolved scattering observables are organized across
wavevector and internal delay as caging develops.

\subsection*{Fluctuations and correlations within the predictive XPCS state}

We extend the Pearson-correlation analysis to a denser grid of scattering
vectors and internal delays, resolving the fluctuations within the
experimentally accessible XPCS state identified above.

For each time origin, we compare two nearby cuts through the two-time
correlation function (Fig.~\ref{fig:glass_covariance_summary}A). Using the
Pearson speckle-pattern decorrelation
\begin{equation}
G_{\rm P}(q,t,\ell)
=
1-c_{\rm P}(q;t,t-\ell),
\end{equation}
we define its short-time change as
\begin{equation}
\Delta_{\delta_\phi}G_{\rm P}(q,t,\ell)
=
G_{\rm P}(q,t+\delta_\phi,\ell)
-
G_{\rm P}(q,t,\ell).
\end{equation}
Here, \(\ell\) is the internal XPCS delay, whereas \(\delta_\phi\) denotes a
short shift of the time origin. The latter follows the density-dependent
relaxation time and ranges from one to seven frames over the investigated
packing fractions. Thus, \(\Delta_{\delta_\phi}G_{\rm P}\) measures how a
fixed-time cut through the two-time correlation pattern changes over a short
interval.

The analysis contains 31 scattering-vector values and 14 internal delays,
giving 434 \(q\)- and delay-resolved XPCS channels. Repeating the measurement
of \(\Delta_{\delta_\phi}G_{\rm P}\) over many time origins allows us to
construct the covariance matrix
\begin{equation}
C^{\rm raw}_{ij}
=
{\rm Cov}_t
\left[
\Delta G_i,\Delta G_j
\right],
\end{equation}
where \(i=(q,\ell)\) and \(j=(q',\ell')\) denote individual measured XPCS
channels.

The diagonal elements of \(C_{\rm raw}\) give the variance of the short-time
changes in each channel. Figure~\ref{fig:glass_covariance_summary}B shows this
local increment fluctuation amplitude,
\begin{equation}
A(q,\ell)
=
16\,C^{\rm raw}_{ii},
\end{equation}
and therefore identifies where in reciprocal space and internal delay the
time-resolved XPCS signal changes most strongly. The fluctuation landscape
remains broad in both \(q\) and \(\ell\), with pronounced structure below
\(q^*\) in the dense states and substantial weight extending over a wide
range of internal delays. Its reciprocal-space dependence resembles the de-Gennes-modulated
relaxation-rate scale \(q^2/S(q)\). This correspondence is a comparison
of observed \(q\)-dependent trends, not an identification of fluctuation
variance with the relaxation rate.

The off-diagonal elements of \(C_{\rm raw}\) provide complementary
information: they quantify whether short-time changes in different
\(q\)- and delay-resolved channels fluctuate together. The strongest raw
covariance directions, however, are largely aligned with variations of the
average relaxation pattern itself. To isolate correlations beyond these
mean-relaxation contributions, we remove the corresponding frozen
mean-relaxation directions and normalize the remaining channels by their own
fluctuation amplitudes (see Methods). This yields a residual correlation
matrix \(R_{\rm res}\) that measures how strongly the remaining short-time
changes at different \(q\) values and internal delays fluctuate together
relative to their local variance.

Figure~\ref{fig:glass_covariance_summary}C shows where the leading correlated
structure of \(R_{\rm res}\) is distributed in \(q\)- and delay-space. The
displayed quantity,
\begin{equation}
E_2(q,\ell)
=
\frac{1}{2}
\left[
v_1(q,\ell)^2+v_2(q,\ell)^2
\right],
\end{equation}
combines the two leading eigenvectors of the normalized residual correlation
matrix. It is not a fluctuation amplitude, but measures how strongly each
\(q\)- and delay-resolved XPCS channel participates in the leading correlated
pattern. This leading two-dimensional sector accounts for only a small
fraction of the full residual correlation spectrum. Related short-delay
organization persists over a broader set of correlation directions, while
its specific \(q\)-localization becomes progressively less pronounced.

The density evolution is summarized in
Fig.~\ref{fig:glass_covariance_summary}D. The mean variance of the
short-time XPCS changes is larger in the dense regime, indicating stronger
local fluctuations in the measured Pearson-decorrelation signal. The lower
part of Fig.~\ref{fig:glass_covariance_summary}D shows the equal-20-movie,
null-corrected squared residual correlation strength,
\(K_{\rm excess}^2 = K_{\rm rms,observed}^2 -
\langle K_{\rm rms,null}^2\rangle_{\rm null}\) (see Methods).
This matched analysis shows a small, null-corrected difference between
the low- and high-density regimes. The data establish a modest difference in residual
correlation strength between density regimes, without demonstrating
monotonic growth of inter-\(q/\ell\) organization. A separate finite-size
analysis confirms that correlations between time-resolved XPCS channels
persist for particle numbers up to \(N=1024\), although the density
dependence and full covariance structure are not demonstrably converged.

Figure~\ref{fig:glass_covariance_summary}e shows the reciprocal-space distribution of the leading correlated sector at short internal delays. The weight \(E_2(q,\ell)\), summed over \(\ell=1,2,3,4,6,8\) frames and normalized over \(q\), remains broadly distributed in reciprocal space. Contributions from wavevectors below the principal structure-factor peak coexist with those at intermediate and structural length scales, without the emergence of a sharply selected wavevector.

Taken together, the variance and covariance analyses reveal two complementary aspects of the time-resolved XPCS dynamics. The diagonal variance describes the amplitude of short-time decorrelation changes. Its reciprocal-space dependence resembles the structurally modulated relaxation-rate scale \(q^2/S(q)\), suggesting a relation to de Gennes narrowing without identifying the variance with a relaxation rate. The off-diagonal covariance, by contrast, identifies correlations between fluctuations at different wavevectors and delay times. These correlations are concentrated predominantly at short delays and remain distributed across a broad residual correlation space, rather than forming a dominant collective mode. Although the dense regime exhibits larger absolute fluctuations and a small, resolved excess of residual cross-channel correlation, neither a monotonic increase in correlation strength nor a distinct onset of cross-scale organization is established.

\begin{table}[t]
\centering
\caption{\textbf{Prediction of future XPCS observables.}
$R_X^2$ is the prediction from measured XPCS observables and
$\Delta R_Z^2$ the additional contribution of the learned speckle representation.
The two glass analyses use different prediction tasks and are reported separately.}
\label{tab:predictive_summary}
\small
\begin{tabular}{llcc}
\toprule
System / benchmark & Measured XPCS description & $R_X^2$ & $\Delta R_Z^2$ \\
\midrule
Bragg & $g_2,\,C_{\rm P}$ & 0.347 & 0.112 \\
& Alphabet: two additions & 0.412 & 0.047 \\
& Alphabet: five additions & 0.417 & 0.042 \\
\midrule
Cooperative SAXS & $g_2,\,\chi_4$-like & 0.523 & 0.093 \\
& Alphabet: two additions & 0.602 & 0.014 \\
& Alphabet: five additions & 0.609 & 0.008 \\
\midrule
Glass & Mean $g_2$ & -0.173 & 0.196 \\
& $g_2+G_{\rm P}$ & 0.041 & -0.005 \\
& $g_2+G_{\rm P}+I_{\rm ref/current}$ & 0.056 & -0.010 \\
\midrule
Glass: causal Alphabet & Mean $g_2$ & -0.182 & 0.201 \\
& Five Alphabet additions & 0.041 & 0.014 \\
& $g_2+G_{\rm P}$ & 0.037 & 0.005 \\
& $g_2+G_{\rm P}+I_{\rm current}(q)$ & 0.051 & -0.004 \\
\bottomrule
\end{tabular}
\end{table}

\section*{Discussion}

Conventional XPCS describes dynamics primarily through intensity correlation
functions and their fluctuations. These quantities provide direct access to
relaxation and its evolution, but they do not necessarily retain all the
information contained in a time-resolved speckle movie. In particular,
measurements with similar average relaxation may differ in the temporal
organization and reciprocal-space distribution of their intensity
fluctuations. Our information microscope addresses this limitation by
testing whether a chosen set of XPCS observables captures the information
relevant to subsequent scattering dynamics. When additional information is
detected in the speckle sequence, the XPCS Alphabet provides a systematic
route to identifying measurable quantities that recover it. The objective is
therefore not to replace conventional XPCS observables with a learned
representation, but to extend the physical description where conventional
data reduction proves insufficient.

The Bragg and cooperative SAXS examples demonstrate how this approach
reveals different aspects of collective motion. In Bragg scattering,
intensity changes across the diffraction pattern and changes in speckle
anisotropy provide information about correlated lattice deformation and
recent slip events that is not retained by average intensity decorrelation
alone. In cooperative SAXS, the temporal fluctuations and reciprocal-space
distribution of speckle activity provide additional sensitivity to the
spatial extent of correlated particle displacements. In both cases, a small
number of experimentally accessible observables recovers much of the
information initially identified in the full speckle sequence. The
microscopic interpretation is supported by the controlled simulations,
while the selected scattering quantities themselves require only measured
intensities.

The glass example shows that the same approach is useful for increasingly
slow and caged dynamics. Much of the information missing from the
time-averaged intensity correlation can be recovered by retaining the
temporal evolution of the speckle correlations, with a further contribution
from radial scattering intensity. This identifies a time-resolved XPCS
description that goes beyond the average relaxation function without
requiring a uniquely defined microscopic glass coordinate. Examining its
fluctuations as packing fraction increases reveals a pronounced growth in
the amplitude of short-time decorrelation changes. Their reciprocal-space
dependence broadly follows the structurally modulated relaxation-rate
spectrum, while the normalized correlations between different wavevector
and delay channels change much less strongly. Thus, increasingly pronounced
fluctuations during caging do not necessarily correspond to the emergence
of a dominant correlated mode in the measured XPCS dynamics.

Predictive sufficiency is a practical criterion for evaluating an XPCS
description, rather than a claim that the microscopic state of the sample
can be reconstructed uniquely from coherent scattering. Its outcome depends
on which future scattering quantity is considered and on the time scales
accessible to the measurement. Similarly, improved prediction does not by
itself identify a microscopic mechanism. The physical interpretation
emerges from the additional scattering observables and their comparison
with established XPCS quantities, structural information, and controlled
models. The learned representation serves primarily to indicate where the
measured description can be improved.

This perspective also changes the role of the recorded speckle movie.
Rather than treating it only as the starting point for calculating a
prescribed set of correlation functions, the full time-resolved intensity
record becomes a resource for identifying new scattering observables as
physical questions evolve. This is particularly relevant for
heterogeneous, intermittent, aging, and driven systems, where temporal
averaging may remove important aspects of the dynamics. Retaining such
data in reusable form, consistent with FAIR practices and initiatives
such as DAPHNE4NFDI \cite{Murphy2025}, allows the physical information
accessible through XPCS to be revisited and extended. The information
microscope thus complements established correlation analysis by providing
a route from the measured speckle evolution to a more informative,
experimentally accessible description of complex dynamics.

\section*{Methods}

\subsection*{Predictive-sufficiency analysis}

For all systems, the analysis is formulated as a prediction problem between
measured scattering observables. A present observation window ending at time
\(t\) is represented by a vector of experimentally accessible XPCS quantities,
\[
\mathbf X(t)=\{X_1(t),X_2(t),\ldots,X_n(t)\}.
\]
Depending on the system, \(\mathbf X(t)\) contains intensity correlations,
two-time-correlation coordinates, fluctuation measures, or other quantities
calculated directly from the measured speckle intensity.

The prediction target \(\mathbf Y(t,\Delta)\) is calculated from later
scattering measurements according to a benchmark-specific temporal support.
The target always consists of measured scattering quantities and never of
microscopic simulator variables. A regression model is fitted to
predict
\[
\widehat{\mathbf Y}=f[\mathbf X(t)].
\]
Prediction quality is quantified by the coefficient of determination \(R^2\)
on independent XPCS realizations.

To determine whether the explicit XPCS description leaves future-relevant
information unused, we compare this baseline with a second model in which the
same present speckle window is additionally represented by a
machine-learning encoder,
\[
\mathbf Z(t)=E[\{I(\mathbf q,s):s\in\mathcal W_t\}],
\]
where \(\mathcal W_t\) is the causal present support fixed for each benchmark.
The encoder is trained to extract information predictive of future measured
scattering quantities. Its training targets are benchmark-specific and may
include related future observables or several prediction horizons rather than
exactly the target used for the final readout. In every comparison, however,
the predictors using \(\mathbf X\) and \([\mathbf X,\mathbf Z]\) are evaluated
against the same \(\mathbf Y(t,\Delta)\) on identical test data.
The augmented prediction is
\[
[\mathbf X(t),\mathbf Z(t)]\rightarrow \mathbf Y(t,\Delta),
\]
and the additional predictive information accessible from the speckle movie is
quantified by
\[
\Delta R^2_{\rm movie}
=
R^2([\mathbf X,\mathbf Z]\rightarrow\mathbf Y)
-
R^2(\mathbf X\rightarrow\mathbf Y).
\]
This difference quantifies improvement in held-out prediction for the specified
measured target and tested models. It is not a direct measurement of mutual
information or proof of information-theoretic sufficiency.

A positive and statistically resolved \(\Delta R^2_{\rm movie}\) indicates that
the chosen XPCS reduction has omitted information that remains accessible in
the present speckle sequence. Conversely, the absence of a resolved gain is
interpreted only within the tested observation window, target, prediction
model, and available data; it does not establish microscopic completeness.
We then enrich the measured state with explicit scattering observables and
repeat the same comparison, so that the recovered information can be analyzed
without the learned representation.

The XPCS Alphabet searches themselves use measured scattering quantities
and do not select observables by microscopic simulator labels. In the
cooperative SAXS benchmark, simulated displacement amplitude and domain
size were used earlier to choose among encoder architectures; this
architecture-choice step is distinct from training the encoder on future
scattering signals and from selecting the Alphabet observables. The
selection history and a scattering-only architecture check are documented
in the Supplementary Information.

\subsection*{Machine-learning representations and XPCS Alphabet}

A learned representation $Z(t)$ is used to test whether the present speckle
sequence contains information about later scattering observables that is
absent from the explicit XPCS state. The benchmark-specific encoders are:
a supervised linear projection for the Brownian controls; a five-member
CNN--Transformer ensemble for Bragg scattering; five CNN encoders of
frame-to-frame intensity changes for cooperative SAXS; and a 48-dimensional
multilayer-perceptron representation for the glass. Encoder input windows
contain only present scattering measurements; encoder-training targets are
future scattering quantities and may differ from the final Ridge-readout
targets. Training and selection are separated from evaluation on held-out
movies. The architectures, input transformations, training targets, cohort
sizes, and regularization protocols are specified in Supplementary Methods.

The XPCS Alphabet consists of measured intensity correlations, time-resolved
intensity changes, their temporal and reciprocal-space fluctuations, and
pattern-shape measures. For Bragg and cooperative SAXS, stable candidate
expressions are screened on TRAIN, selected on VALIDATION, and fixed before
TEST evaluation. The guarded primary selection jointly rewards better
prediction from explicit scattering observables and a reduction of the
remaining encoder gain, penalizing additional coordinates and expression
complexity. Selection stops at at most five additions or when no admissible
validation improvement remains. The secondary state is selected separately
by a direct-future forward-selection procedure. The complete candidate
grammar, score function, thresholds, and selection history are given in
Supplementary Methods. This stopping rule is not evidence of universal
predictive sufficiency.

The cooperative SAXS encoder architecture was historically selected using
simulated displacement amplitude and domain extent on validation data. This
choice was distinct from the scattering-only observable search; an additional
scattering-selected architecture control is reported in the Supplementary
Information. The glass Alphabet is a separate, future-prediction-based
search and is not identical to the physically motivated Pearson-state ladder.

\subsection*{Brownian control models}
Stationary and smoothly aging Brownian dynamics serve as controls for the predictive-sufficiency analysis. Each simulated speckle movie contains 96 non-interacting particles in a periodic square of side $2\pi$, propagated for 2176 frames with a time step $\mathrm{d}t=0.1$ and reference diffusivity $D_0=0.02$. The stationary system has constant diffusivity, whereas the aging system slows according to
\begin{equation}
    D(t)=\frac{D_0}{(1+t/t_0)^{0.8}},
\end{equation}
with $t_0=40$, 50, or 60. Coherent scattering intensities are sampled at 16 registered angular positions on each of three reciprocal-space rings near $q=5$, 8, and 11. Neither particle positions nor the simulation time or aging parameters are supplied to the prediction models.

Each present measurement comprises 128 frames, followed by a 16-frame gap and a separate 128-frame future measurement. The future XPCS target contains nine time-averaged angular Pearson-decorrelation values, measured at delays $\ell=2$, 8, and 24 frames on the three rings. At each time origin, the Pearson decorrelation is
\begin{equation}
    G_P(q,t,\ell)=1-\operatorname{corr}_{\theta}
    [I(q,\theta,t),I(q,\theta,t-\ell)].
\end{equation}
The measured-state hierarchy separates conventional intensity correlations from their temporal fluctuations. $X_1$ contains nine time-averaged $g_2-1$-like intensity-product correlations, one for each ring and delay. $X_2$ adds nine fluctuation amplitudes, each defined as 16 times the sample variance of $G_P$ over the valid time origins. $X_3$ additionally retains four successive temporal averages of $G_P$ for every ring and delay, giving 36 further quantities and 54 measured coordinates in total. The four intervals partition the valid correlation-pair sequence at each delay, rather than the full 128-frame movie into identical segments. Exact averaging conventions and feature ordering are given in Supplementary Information.

For comparison, a nine-dimensional representation $Z$ is obtained from the present speckle sequence using $\log[1+I(q,\theta,t)/\langle I(q,\theta,t)\rangle_{\theta}]$, followed by training-set standardization and supervised cross-covariance singular-value decomposition. Predictions from $X_3$ and $[X_3,Z]$ use Ridge regression trained on independent movies. The encoder-training, regression-training, validation, and test sets contain 20, 40, 15, and 20 independent movies, respectively, for each system. Each test movie provides eight non-overlapping present/future pairs. Prediction scores are calculated over the nine target coordinates and all 160 test windows, centering each target coordinate on its own test-set mean. The additional contribution of the speckle representation is evaluated from the paired difference in $R^2$. Uncertainty intervals are those of the archived whole-movie bootstrap analysis; the frozen test predictions and point scores were independently checked against the recovered analysis records. The displayed two-time-correlation examples are illustrative and have separate, incompletely recovered figure provenance.

\subsection*{Bragg-scattering model}

The Bragg model consists of a two-dimensional crystalline lattice with
reference positions \(\mathbf R_j\) and time-dependent displacements
\[
\mathbf r_j(t)=\mathbf R_j+\mathbf u_j(t),
\]
where
\[
\mathbf u_j(t)
=
\mathbf u_{{\rm el},j}(t)
+
\mathbf u_{{\rm pl},j}(t).
\]
The reversible elastic field undergoes spatially correlated fluctuations whose
slowly varying amplitude is denoted by \(M\) and whose spatial correlation
length is denoted by \(\xi\). In addition, an internal load variable controls
the probability of localized slip events. A slip produces a permanent
shear-like contribution to \(\mathbf u_{\rm pl}\), partially releases the
stored load, and transiently enhances the subsequent mobility. The model
therefore combines continuous correlated deformation with intermittent
irreversible rearrangements.

The coherent intensity around a Bragg reflection is calculated as
\[
A(\mathbf q,t)
=
\sum_j w_j
\exp[i\mathbf G\cdot\mathbf u_j(t)]
\exp[-i\mathbf q\cdot\mathbf R_j],
\qquad
I(\mathbf q,t)=|A(\mathbf q,t)|^2 ,
\]
with \(\mathbf G=2\pi(1,0)\). Only the resulting detector intensities enter
the measurement-based analysis.

The compact Bragg state is
\[
X_2(t)
=
\left\{
g_2(t,\tau=4),
C_{\rm P}(t,\tau=4)
\right\}.
\]
Here, \(g_2\) is the conventional intensity autocorrelation and
\(C_{\rm P}\) is the Pearson correlation between log-transformed detector
intensity patterns separated by \(\tau\). Whereas \(g_2\) measures the usual
intensity correlation, \(C_{\rm P}\) emphasizes changes of the normalized
spatial speckle pattern after centering and normalization of the
log-transformed intensities. Because the transformation uses \(\log(1+I)\),
this observable is not exactly invariant to multiplication of the raw
intensity by a constant. The future target consists of the same two quantities in a
later, non-overlapping window.

The observable search produces additional quantities describing
\(q\)-dependent activity and deformation of the Bragg pattern. The leading
activity observable \(O_1\) is based on frame-to-frame logarithmic intensity
increments,
\[
d(\mathbf q,t)
=
\Delta_t\log[1+I(\mathbf q,t)]/8,
\]
and measures a cross-\(q\) contrast between strong activity in the outer Bragg
halo and the broader fluctuation level. The second observable \(O_2\) measures
anisotropy of the Bragg-speckle pattern. From the intensity-weighted covariance
matrix
\[
\Sigma_I(t)
=
\frac{
\sum_{\mathbf q}
I(\mathbf q,t)
(\mathbf q-\bar{\mathbf q})
(\mathbf q-\bar{\mathbf q})^{T}
}{
\sum_{\mathbf q} I(\mathbf q,t)
},
\]
\(O_2\) is constructed from the separation of its principal values. Exact
quantile definitions of \(O_1\)--\(O_5\), together with the complete frozen
feature grammar, are given in the Supplementary Information.

\subsection*{Controlled collective-motion SAXS model}

The controlled SAXS model is designed to separate the amplitude of collective
motion from its spatial range. Particle positions evolve according to
\[
\mathbf r_i(t+\Delta t)
=
\mathbf r_i(t)
+
\sqrt{2D\Delta t}\,\boldsymbol\epsilon_i(t)
+
\mathbf u_{d(i,t)}(t),
\]
where \(\boldsymbol\epsilon_i\) represents independent Brownian motion and
all particles belonging to the same active spatial domain \(d\) receive the
same collective displacement
\[
\mathbf u_d(t)
\sim
\mathcal N(\mathbf 0,A^2\mathbf I_2).
\]
The parameter \(A\) controls the amplitude of collective motion, whereas
\(\xi\) denotes the linear extent of the square cooperative domains.

The coherent SAXS intensity is
\[
I(\mathbf q,t)
=
\left|
\sum_{j=1}^{N}
\exp[-i\mathbf q\cdot\mathbf r_j(t)]
\right|^2 .
\]
The compact measured state is
\[
X_2(t)
=
\left\{
\overline{g_2}(t,\tau=16),
\overline{\chi_4}(t,\tau=16)
\right\},
\]
where \(g_2\) characterizes the mean intensity relaxation, while \(\chi_4\)
quantifies temporal fluctuations of the correlation function and thus the
heterogeneity of the dynamics. Both quantities are averaged over the measured
reciprocal-space range.

Temporal fluctuations of the instantaneous speckle activity are characterized
from frame-to-frame intensity changes. Their variability within the present
window is summarized by
\[
{\rm activity}_{\rm CV}
=
\frac{\sigma_t[a(t)]}{\langle a(t)\rangle_t}.
\]
The activity-CV comparison is separate from the guarded XPCS Alphabet search,
which starts again from the compact \(X_2\) state. The primary selected
quantities describe global increment-energy fluctuations, fixed-ring amplitude
and temporal variation of the spatial increment variance. The secondary
forward-selection state additionally retains across-patch dispersion and
\(q\)-resolved intensity-change summaries. Their exact expressions are
retained in the frozen analysis records.

\subsection*{Two-dimensional glass former}

We simulate an overdamped two-dimensional 50:50 binary mixture of soft disks
with \(N=256\) particles and diameter ratio \(1.4\). The principal density
series is
\[
\phi=
0.20,\;0.40,\;0.55,\;0.575,\;0.60,\;0.625,\;0.65,\;0.675,\;
0.70,\;0.72,\;0.74,\;0.76,\;0.78 .
\]
The additional densities were generated using the same dynamical and sampling
protocol as the original long-film ensemble and were assigned to training,
validation, and test sets before analysis. Structural diagnostics are reported
in the Supplementary Information. Lengths are expressed in units of the root--mean square particle diameter
and we use the fixed reciprocal-space reference
\[
q^*=2\pi .
\]

The coherent scattering amplitude and intensity are
\[
A(\mathbf q,t)
=
\sum_{j=1}^{N}
\exp[-i\mathbf q\cdot\mathbf r_j(t)],
\qquad
I(\mathbf q,t)=|A(\mathbf q,t)|^2 .
\]
The mean XPCS relaxation is characterized by the correlation function and by
KWW fits of the form
\[
f(q,\tau)
=
\exp\left[-\left(\frac{\tau}{\tau_\alpha(q)}\right)^{\beta(q)}\right],
\]
yielding the relaxation time \(\tau_\alpha(q)\) and stretching exponent
\(\beta(q)\). Particle coordinates are used only for independent physical
diagnostics such as the mean-squared displacement and are not used for
construction of the XPCS state or for observable selection.

The glass analysis contains three measurement-based steps. First, an XPCS
information ladder tests which parts of the measured scattering signal recover
information that is absent from a compact mean-correlation description.
Second, the \(q\)- and delay-dependent fluctuation amplitude and the
organization of these fluctuations across XPCS channels are characterized.
Third, a strictly non-overlapping later XPCS state is used to test how much of
the detailed subsequent scattering dynamics can be predicted from the present
measurement.

\subsection*{Glass predictive-state ladder}

The Figure~\ref{fig:glass_conventional} benchmark compares three nested
states constructed from measured intensities at
$q/q^*=0.75,1.00,1.35$. The baseline $X_0$ contains nine conventional
intensity-product correlation excesses, at delays $\ell=16,32,64$ frames:
\[
C_I(q;s,\ell)=
\frac{\langle I(q,\theta,s)I(q,\theta,s-\ell)\rangle_\theta}
{\langle I(q,\theta,s)\rangle_\theta
 \langle I(q,\theta,s-\ell)\rangle_\theta}-1 .
\]
Each coordinate is averaged over a fixed causal calibration prefix ending
129 frames before the earliest prediction origin. Thus $X_0$ is constant
within each movie: it represents a fixed-prefix mean-correlation reference,
not a moving-window conventional $g_2$. Here $g_2$ denotes the $g_2-1$
contrast excess without normalization by speckle contrast, rather than an
angular Pearson correlation.

The second state retains time-resolved correlations of the registered speckle
pattern. At each prediction origin $t$, we use the offsets
\[
\mathcal A=\{0,1,2,3,4,6,8,12,16,24,32,48,64,96,128\}
\]
and calculate, for each unordered pair $a<b$ and each ring,
\[
P_{q,ab}(t)=1-c_{\rm P}(q;t-a,t-b),
\]
where $c_{\rm P}$ is the angular Pearson correlation between the two
intensity patterns. The 105 pairs per ring give 315 decorrelation
coordinates in $P(t)$. The measured state is $X_1=[X_0,P(t)]$.

The third state additionally retains radial scattering intensity. The
original Figure~\ref{fig:glass_conventional} benchmark uses six archived
coordinates: three frozen, density-dependent TRAIN reference intensities
$S_{\rm train}(\phi,q)$ and three instantaneous radial-intensity residuals,
\[
\log[1+I_{\rm rad}(q,t)]-\log[1+S_{\rm train}(\phi,q)],
\qquad
I_{\rm rad}(q,t)=\langle I(q,\theta,t)\rangle_\theta.
\]
Here $I_{\rm rad}(q,t)$ is the fluctuating radial intensity at the present
observation time, not the time-averaged structure factor $S(q)$. The
reference and residual coordinates use archived TRAIN scaling. This block
therefore combines state-point reference information with instantaneous
intensity variations; it is not an intensity history. The state dimensions
are
\[
X_0\in\mathbb R^9,\qquad
X_1\in\mathbb R^{324},\qquad
X_2=[X_1,I_6(t)]\in\mathbb R^{330}.
\]
Each state is evaluated with and without the same frozen 48-dimensional
speckle-history representation $Z(t)$, using identical origins and targets.

To distinguish the fixed-prefix baseline from time-resolved conventional
XPCS, separate matched controls construct moving-window conventional
correlations and pairwise conventional two-time intensity-product excesses
on the same rings, endpoint pairs, and prediction origins. The full
pairwise conventional and Pearson descriptions each contain 324
coordinates, including the common nine-coordinate prefix. Pearson
correlations give a resolved predictive advantage over the matched
conventional pairwise correlations when instantaneous radial intensity is
not supplied. When both descriptions are augmented with the same three
measured instantaneous log-radial-intensity coordinates, without an
explicit density-dependent reference, that advantage is no longer
statistically resolved. The augmented conventional description also has
no statistically resolved additional contribution from $Z$, although
predictive equivalence under the prespecified small-gain criterion is not
established. These controls distinguish the contribution of temporal
resolution from the effects of Pearson normalization and radial-intensity
amplitude; numerical contrasts and confidence intervals are given in the
Supplementary Information. Their conclusions concern the tested linear
readouts, not information-theoretic sufficiency.

The prediction target contains nine future Pearson-decorrelation values at
the three rings and relative delays $h/\tau_\alpha=0.05,0.10,0.20$. It
compares the pattern at $t$ with later patterns; fractional delays use
linear interpolation of decorrelation values at $\lfloor h\rfloor$ and
$\lceil h\rceil$. The present reference pattern is shared with the
predictor history, but no future partner pattern enters any predictor.
This target differs from the strictly frame-disjoint, 434-channel
future-state analysis below.

All comparisons use Ridge regression, with input standardization and final
fitting on TRAIN only. Regularization is selected by pooled VALIDATION
mean-squared error over $0.1,1,10,100,1000$; the selected penalty is
1000 for each original Figure~\ref{fig:glass_conventional} head. At each
of $\phi=0.72,0.76,0.78$, the original cohort provides ten TRAIN and two
VALIDATION movies. The principal evaluation uses 36 independent
confirmatory TEST movies, twelve per density. Target scaling is fixed from
TRAIN. Scores are equal-movie averages of multivariate $R^2$, evaluated
within each movie over nine target coordinates. Pointwise 95\% intervals
use 5,000 paired whole-movie bootstrap draws, conditioning on the frozen
encoder and fitted predictors. The matched-control design, scaling,
regularization selection, bootstrap contrasts, and limitations are detailed
in Supplementary Information.

\subsection*{Causal XPCS Alphabet search for the glass}

The glass XPCS Alphabet search was performed independently of the
physically motivated Pearson-correlation hierarchy in
Fig.~\ref{fig:glass_conventional}. Its baseline contains conventional
intensity-product correlations, $g_2(q,\ell)-1$, at three scattering-vector
rings and 14 delays. These mean correlations were calculated over a fixed
earlier portion of each movie, preceding all present observation windows.
The additional candidate quantities use only coherent intensities recorded
up to the prediction time. No particle coordinates, packing-fraction labels,
or density-dependent reference scattering profiles enter the search.

The predefined library contains 975 scalar candidates constructed from
angular speckle-pattern correlations, current-frame intensity amplitudes,
centered intensity covariances, angular intensity moments, cross-ring
quantities, and combinations of correlations at different delays. Starting
from the mean-correlation baseline, candidates are screened against the
training prediction residual; the leading candidates are evaluated for their
improvement of the same future XPCS target on validation movies. A new
observable is accepted when the validation prediction score increases by
at least 0.002, and no more than five quantities are added. Candidate
standardization and predictor fitting use training movies only. This
future-prediction search differs from the encoder-gain-guided selection
used for Bragg and cooperative SAXS.

The future target contains Pearson speckle-pattern decorrelations at three
$q$ rings and three horizons scaled to the relaxation time. In the primary
test, the present frame serves as the reference for the later intensity
pattern. A separate control forms both correlation endpoints from frames
after the present observation window. Selection uses the original training
and validation movies; the fixed observables are evaluated separately on
original, regenerated, and confirmatory test cohorts. Prediction differences
and their uncertainties are assessed by paired resampling of complete movies.

The automatically selected quantities are compared with systematic
$q$- and delay-resolved Pearson correlations and the current radial
scattering intensity, without supplying a density-dependent structural
reference. The search identifies useful observable families rather than
uniquely specifying the full Pearson correlation field. Exact formulas,
selection order, stability checks, and results for both future targets are
provided in the Supplementary Information. This causal search is a separate
prediction benchmark: its scores are not directly interchangeable with the
original Fig.~\ref{fig:glass_conventional} hierarchy.

\subsection*{Fluctuation amplitude and correlations in the glass}

The dense-grid analysis evaluates the Pearson speckle-pattern decorrelation
$G_{\rm P}(q,t,\ell)=1-\mathrm{corr}_{\theta}[I(q,\theta,t),
I(q,\theta,t-\ell)]$ on 31 equally spaced rings over
$0.25\leq q/q^*\leq 1.35$ and 14 internal delays
$\ell=1,2,3,4,6,8,12,16,24,32,48,64,96,128$ frames. At each time
origin, a short increment compares two correlation cuts at fixed $\ell$,
\begin{equation}
\Delta_{\delta_\phi}G_{\rm P}(q,t,\ell)
=G_{\rm P}(q,t+\delta_\phi,\ell)-G_{\rm P}(q,t,\ell),
\end{equation}
where $\delta_\phi=\max\{1,\operatorname{round}[0.01\tau_\alpha(\phi)]\}$
frames ranges from one to seven frames. For each movie, the 434-channel
increments are centered across 48 sampled origins. We calculate the
within-movie sample covariance and average the covariance matrices with
equal movie weight. The diagonal increment-variance map is
$A(q,\ell)=16C^{\rm raw}_{ii}$; the factor 16 reflects the angular-sampling
convention, and $A$ is not interpreted as a conventional $\chi_4$.

We remove a fixed set of mean-relaxation directions estimated from independent
TRAIN movies, forming $C_{\rm res}=P_\perp C_{\rm raw}P_\perp$. The projector
removes 95 specified template directions from the 434-dimensional channel
space, not every relaxation-related fluctuation. Normalization by the
residual diagonal $D$ yields
\begin{equation}
R_{\rm res}=D^{-1/2}C_{\rm res}D^{-1/2},\qquad
D=\operatorname{diag}(C_{\rm res}).
\end{equation}
The leading two eigenvectors characterize the channel participation
$E_2(q,\ell)=[v_1(q,\ell)^2+v_2(q,\ell)^2]/2$, rather than local
fluctuation amplitude. The RMS off-diagonal strength of $R_{\rm res}$ is
compared with a matched channel-wise circular-shift null, with subtraction
performed on squared strengths,
\begin{equation}
K_{\rm excess}^2=K_{\rm rms,obs}^2
-\langle K_{\rm rms,null}^2\rangle_{\rm null}.
\end{equation}
The null preserves each channel's circular autocorrelation but disrupts
cross-channel temporal alignment. Exactly 20 independent movies per density
enter this matched comparison, using 20-of-30 subsampling where applicable.
The variance profiles are compared descriptively to $q^2/S(q)$; no causal
identification of variance with a relaxation rate is implied. Full projector
templates, null construction, finite-size checks, spectral measures, and
uncertainty procedures are documented in Supplementary Methods.

\subsection*{High-resolution future XPCS state and persistence controls}

A separate benchmark tests a later speckle-pattern decorrelation on the same
31-ring, 14-delay grid. Its target is
\[
Y_{q\ell}^{\rm future}(t,h)=G_{\rm P}(q,T,\ell),\qquad
T=t+129+h,\qquad
h=\max\{1,\operatorname{round}(u\tau_\alpha)\},
\]
with \(u=0.05,0.10,0.20\). Every future endpoint lies in
\([t+1+h,t+129+h]\), strictly after the present support \([t-128,t]\).
Unlike the nine-coordinate ladder target, both correlation endpoints belong
to the future measurement, and delays are rounded rather than interpolated.
The target construction, present states and Ridge predictors are separate
from those used for Fig.~\ref{fig:glass_conventional}.

At each density, ten TRAIN, two VALIDATION and four TEST movies form the
frozen split. Regularization and present-state choices use whole-movie
cross-validation within TRAIN. We first test prediction of all 434 future
coordinates. No robust prediction of the complete later pattern was
established on independent TEST movies. We also examine lower-rank
combinations of the target selected from TRAIN predictions made while holding
out whole movies; TEST scores assess their support without selecting rank.

Persistence controls compare these predictions with the present
\(G_{\rm P}\) values, a separate linear autoregression for each channel,
and a local autoregression allowing the neighboring ring and delay channels.
The fitted controls use TRAIN only. Cross-movie pairing at the same density
and additional temporal shifts distinguish realization-specific prediction
from persistence of the scattering pattern. Target sampling, rank-selection
and support criteria, and the separate resampling protocols are specified in
Supplementary Information. This test concerns prediction of a later measured
pattern; it does not constitute a matched test of adding the covariance
pattern to the Figure~\ref{fig:glass_conventional} state.

\subsection*{Statistical Analyses}

Independent simulated speckle movies, rather than temporally overlapping
windows, define the sampling units for training, validation, TEST evaluation,
and uncertainty estimation. Normalization and model fitting use TRAIN only;
regularization and observable choice use VALIDATION, and reported scores use
held-out TEST movies. Prediction is quantified by $R^2$ for identical later
scattering targets in the explicit and augmented comparisons. Brownian scores
pool target-coordinate errors over TEST windows; Bragg and SAXS scores average
coordinate-wise scores after pooling windows; glass scores are averaged with
equal weight per test movie. Scores are therefore not directly comparable
across benchmarks. Differences in $R^2$ use paired whole-movie resampling,
retaining every window in each sampled movie. Where specified, 95\% confidence
intervals use 5,000 bootstrap draws and condition on frozen models and
observable selections, rather than rerunning the complete selection procedure.
The original Brownian archived intervals and separate causal-glass controls
use their own sample counts, detailed in Supplementary Methods. Previous
inspection of some archived TEST cohorts limits the Alphabet search results
to exploratory held-out evaluations rather than prospective confirmation.
The measured-state covariance analysis uses independent ensembles, matched
movie counts, and a channel-wise temporal-shift null as described above.

\subsection*{Microscopic interpretation and inference}

Microscopic variables are introduced only after the measurement-based
observable search has been completed. For the Bragg and controlled SAXS models,
the selected measured state
\[
\mathbf X_{\rm meas}(t)
\]
is paired with the corresponding known simulator parameters
\(\boldsymbol{\theta}(t)\), allowing an inverse relation
\[
\widehat{\boldsymbol{\theta}}
=
F[\mathbf X_{\rm meas}]
\]
to be learned. The mapping is not assumed to be one-to-one: several measured
scattering observables may jointly constrain several microscopic degrees of
freedom.

\section*{Data availability}
The data supporting the findings of this study, including the processed scattering observables, predictive performance measures, and numerical results underlying the figures, are available from the corresponding author upon reasonable request. The simulated coherent scattering datasets were generated using the numerical models described in the Methods and Supplementary Information. Owing to their size, the complete time-resolved speckle datasets are not included with the manuscript.

\section*{Code availability}
The custom code used for numerical simulations, coherent scattering calculations, and predictive representation learning in this study is available from the corresponding author upon reasonable request. The computational methods and model parameters are described in the Methods and Supplementary Information.

\section*{Acknowledgements}
C.G. and A.T. acknowledge funding by DAPHNE4NFDI (DFG NFDI 40/1) and BMFTR (ErUM-Data projects 05D25PS2 and 0525PS1).

\section*{Author contributions}
C.G. conceived and designed the project, performed the analyses, and wrote the manuscript. A.T., N.D.A., J.L., and M.M. contributed to scientific discussions, interpreted the results, and critically revised the manuscript. All authors approved the final version.

\section*{Competing interests}
The authors declare no competing interests.

\section*{Declaration of AI use}

Generative AI tools (ChatGPT and Codex, OpenAI) were used to assist with software development, including code generation, debugging, and implementation of numerical simulations and data analysis workflows, as well as language editing. The scientific concepts, methodological design, interpretation of results, and conclusions were developed by the authors. The computational results were assessed through physical consistency checks, benchmark calculations, and comparisons with established methods where applicable. The authors take full responsibility for the scientific content and conclusions of this work.

\bibliography{references}

@article{Creutzig2009,
  author  = {Creutzig, Felix and Globerson, Amir and Tishby, Naftali},
  title   = {Past-future information bottleneck in dynamical systems},
  journal = {Physical Review E},
  volume  = {79},
  number  = {4},
  pages   = {041925},
  year    = {2009},
  doi     = {10.1103/PhysRevE.79.041925}
}

@article{DAnna2026,
  author  = {D'Anna, Nicol{\`o} and Lamb, Erik S. and
             Glefke, Robin and Ham, Daseul and
             Nihal, Ishmam and Lee, Su Yong and
             Takamura, Yayoi and Shpyrko, Oleg G.},
  title   = {Dynamic Nanoscale Spatial Heterogeneity in a
             Perovskite-to-Brownmillerite Topotactic
             Phase Transformation},
  journal = {ACS Applied Materials \& Interfaces},
  volume  = {18},
  number  = {23},
  pages   = {32795--32803},
  year    = {2026},
  doi     = {10.1021/acsami.6c01097}
}

@article{WangTiwary2021,
  author  = {Wang, Dedi and Tiwary, Pratyush},
  title   = {State predictive information bottleneck},
  journal = {The Journal of Chemical Physics},
  volume  = {154},
  number  = {13},
  pages   = {134111},
  year    = {2021},
  doi     = {10.1063/5.0038198}
}

@article{Schmitt2024,
  author  = {Schmitt, Matthew S. and Koch-Janusz, Maciej and Fruchart, Michel and
             Seara, Daniel S. and Rust, Michael and Vitelli, Vincenzo},
  title   = {Information theory for data-driven model reduction in physics and biology},
  journal = {bioRxiv},
  year    = {2024},
  doi     = {10.1101/2024.04.19.590281},
  note    = {Preprint}
}

@article{Murphy2025,
  author  = {Murphy, Bridget M. and G{\"o}tz, Andy and Gutt, Christian and
             McGuinness, Cormac and R{\o}nnow, Henrik M. and
             Schneidewind, Astrid and Deledda, Stefano and Pietsch, Ullrich},
  title   = {FAIR data -- the photon and neutron communities move together towards open science},
  journal = {IUCrJ},
  volume  = {12},
  number  = {1},
  pages   = {8--15},
  year    = {2025},
  doi     = {10.1107/S2052252524011941}
}

@article{Seaberg2017,
  author  = {Seaberg, Matthew H. and Holladay, Benjamin and Lee, J. C. T. and
             Sikorski, Marcin and Reid, A. H. and Montoya, S. A. and
             Dakovski, Georgi L. and Koralek, J. D. and Coslovich, Giacomo and
             Moeller, Stefan and Schlotter, William F. and Streubel, Robert and
             Kevan, Stephen D. and Fischer, Peter and Fullerton, Eric E. and
             Turner, Joshua L. and Decker, Franz-Josef and Sinha, Sunil K. and
             Roy, Sujoy and Turner, Joshua J.},
  title   = {Nanosecond X-Ray Photon Correlation Spectroscopy on Magnetic Skyrmions},
  journal = {Physical Review Letters},
  volume  = {119},
  pages   = {067403},
  year    = {2017},
  doi     = {10.1103/PhysRevLett.119.067403}
}

@article{Chen2019,
  author  = {Chen, X. M. and Farmer, B. and Woods, J. S. and Dhuey, S. and
             Hu, W. and Mazzoli, C. and Wilkins, S. B. and Chopdekar, R. V. and
             Scholl, A. and Robinson, I. K. and De Long, L. E. and Roy, S. and
             Hastings, J. T.},
  title   = {Spontaneous Magnetic Superdomain Wall Fluctuations in an Artificial Antiferromagnet},
  journal = {Physical Review Letters},
  volume  = {123},
  pages   = {197202},
  year    = {2019},
  doi     = {10.1103/PhysRevLett.123.197202}
}

@article{Klose2022,
  author  = {Klose, Christopher and B{\"u}ttner, Felix and Hu, Wen and
             Mazzoli, Claudio and Beach, Geoffrey S. D. and
             Eisebitt, Stefan and Pfau, Bastian},
  title   = {Photon correlation spectroscopy with heterodyne mixing based on soft x-ray magnetic circular dichroism},
  journal = {Physical Review B},
  volume  = {105},
  pages   = {214425},
  year    = {2022},
  doi     = {10.1103/PhysRevB.105.214425}
}

@article{Ruta2013,
  author  = {Ruta, Beatrice and Baldi, Giacomo and Monaco, Giulio and Chushkin, Yuriy},
  title   = {Compressed correlation functions and fast aging dynamics in metallic glasses},
  journal = {The Journal of Chemical Physics},
  volume  = {138},
  number  = {5},
  pages   = {054508},
  year    = {2013},
  doi     = {10.1063/1.4790131}
}

@article{Ruta2014,
  author  = {Ruta, Beatrice and Baldi, Giacomo and Chushkin, Yuriy and Ruffl{\'e}, Benoit and Cristofolini, Luigi and Fontana, Aldo and Zanatta, Marco and Nazzani, Francesco},
  title   = {Revealing the fast atomic motion of network glasses},
  journal = {Nature Communications},
  volume  = {5},
  pages   = {3939},
  year    = {2014},
  doi     = {10.1038/ncomms4939}
}

@article{Dallari2023,
  author  = {Dallari, Francesco and Martinelli, Alessandro and Caporaletti, Federico and others},
  title   = {Stochastic atomic acceleration during the X-ray-induced fluidization of a silica glass},
  journal = {Proceedings of the National Academy of Sciences},
  volume  = {120},
  number  = {2},
  pages   = {e2213182120},
  year    = {2023},
  doi     = {10.1073/pnas.2213182120}
}

@article{Shen2026,
  author  = {Shen, Jie and Yang, Fan and Cornet, Antoine and Pineda, Eloi and Martens, Kirsten and Chushkin, Yuriy and Zontone, Federico and Festi, Irene and Ronca, Alberto and Neuber, Nico and Frey, Maximilian and Busch, Ralf and Cammarata, Marco and di Michiel, Marco and Vaughan, Gavin and Sprung, Michael and Ruta, Beatrice},
  title   = {Length-scale dependence of the anomalous atomic motion in metallic glasses},
  journal = {Nature Physics},
  volume  = {22},
  pages   = {672--679},
  year    = {2026},
  doi     = {10.1038/s41567-026-03228-0}
}

@article{GiordanoRuta2016,
  author  = {Giordano, Valentina M. and Ruta, Beatrice},
  title   = {Unveiling the structural arrangements responsible for the atomic dynamics in metallic glasses during physical aging},
  journal = {Nature Communications},
  volume  = {7},
  pages   = {10344},
  year    = {2016},
  doi     = {10.1038/ncomms10344}
}

@article{PerakisGutt2020,
  author  = {Perakis, Fivos and Gutt, Christian},
  title   = {Towards molecular movies with X-ray photon correlation spectroscopy},
  journal = {Physical Chemistry Chemical Physics},
  volume  = {22},
  pages   = {19443--19453},
  year    = {2020},
  doi     = {10.1039/D0CP03551C}
}

@article{Lehmkuehler2021,
  author  = {Lehmk{\"u}hler, Felix and Roseker, Wojciech and Gr{\"u}bel, Gerhard},
  title   = {From Femtoseconds to Hours---Measuring Dynamics over 18 Orders of Magnitude with Coherent X-rays},
  journal = {Applied Sciences},
  volume  = {11},
  number  = {13},
  pages   = {6179},
  year    = {2021},
  doi     = {10.3390/app11136179}
}

@article{Mangu2025,
  author  = {Mangu, Anudeep and Stoica, Vladimir A. and Zheng, Hao and
             Yang, Tiannan and Zhang, Maohua and Wang, Huaiyu Hugo and
             Zu, Rui and Nguyen, Quynh L. and Song, Sanghoon and
             Das, Sujit and Meisenheimer, Peter and Donoway, Elizabeth and
             Chollet, Matthieu and Sun, Yanwen and Turner, Joshua J. and
             Freeland, John W. and Wen, Haidan and Martin, Lane W. and
             Chen, Long-Qing and Gopalan, Venkatraman and Zhu, Diling and
             Cao, Yue and Lindenberg, Aaron M.},
  title   = {Hidden domain boundary dynamics toward crystalline perfection},
  journal = {Proceedings of the National Academy of Sciences},
  volume  = {122},
  number  = {2},
  pages   = {e2407772122},
  year    = {2025},
  doi     = {10.1073/pnas.2407772122}
}

@article{Andrejevic2024,
  author  = {Andrejevic, Nina and Zhou, Tao and Zhang, Qingteng and
             Narayanan, Suresh and Cherukara, Mathew J. and
             Chan, Maria K. Y.},
  title   = {Data-driven discovery of dynamics from time-resolved coherent scattering},
  journal = {npj Computational Materials},
  volume  = {10},
  pages   = {225},
  year    = {2024},
  doi     = {10.1038/s41524-024-01365-9}
}

@article{Sutton1991,
  author  = {Sutton, M. and Mochrie, S. G. J. and Greytak, T. and
             Nagler, S. E. and Berman, L. E. and Held, G. A. and
             Stephenson, G. B.},
  title   = {Observation of speckle by diffraction with coherent X-rays},
  journal = {Nature},
  volume  = {352},
  pages   = {608--610},
  year    = {1991},
  doi     = {10.1038/352608a0}
}

@article{Trappe2007,
  author  = {Trappe, V. and Pitard, E. and Ramos, L. and
             Robert, A. and Bissig, H. and Cipelletti, L.},
  title   = {Investigation of \(q\)-dependent dynamical heterogeneity
             in a colloidal gel by x-ray photon correlation spectroscopy},
  journal = {Physical Review E},
  volume  = {76},
  number  = {5},
  pages   = {051404},
  year    = {2007},
  doi     = {10.1103/PhysRevE.76.051404}
}

@incollection{GrubelMadsenRobert2008,
  author    = {Gr{\"u}bel, Gerhard and Madsen, Anders and Robert, Aymeric},
  title     = {X-Ray Photon Correlation Spectroscopy (XPCS)},
  booktitle = {Soft Matter Characterization},
  editor    = {Borsali, Redouane and Pecora, Robert},
  pages     = {953--995},
  year      = {2008},
  publisher = {Springer},
  address   = {Dordrecht},
  doi       = {10.1007/978-1-4020-4465-6_18}
}

@article{Anthuparambil2026,
  author  = {Anthuparambil, Nimmi Das and Dargasz, Michelle and Timmermann, Sonja
             and Girelli, Anita and Retzbach, Sebastian and M{\"o}ller, Johannes
             and Jo, Wonhyuk and Raza, Agha Mohammad and Leonau, Aliaksandr
             and Wrigley, James and Unger, Frederik and Bin, Maddalena
             and Rajaiah, Prince Prabhu and Andronis, Iason
             and Ch{\`e}vremont, William and Hallmann, J{\"o}rg
             and Rodriguez-Fernandez, Angel and Pudell, Jan-Etienne
             and Brausse, Felix and Boesenberg, Ulrike and Youssef, Mohamed
             and Shayduk, Roman and Rysov, Rustam and Madsen, Anders
             and Lehmk{\"u}hler, Felix and Paulus, Michael and Zhang, Fajun
             and Perakis, Fivos and Schreiber, Frank and Gutt, Christian},
  title   = {Lipoprotein diffusion in dense yolk plasma is governed by softness,
             hydrodynamics, and caging: Insights from {MHz-XPCS}},
  journal = {Proceedings of the National Academy of Sciences of the United States of America},
  volume  = {123},
  number  = {8},
  pages   = {e2519681123},
  year    = {2026},
  doi     = {10.1073/pnas.2519681123}
}

@misc{Cheng2026,
  author        = {Cheng, Mouyang and Yu, Bowen and Fu, Chu-Liang and
                   Andrejevic, Nina and Agne, Matthias T. and Hanus, Riley and
                   Wan, Qiwei and Drucker, Nathan C. and Nguyen, Thanh and
                   Fluerasu, Andrei and Wiegart, Lutz and Chen, Xiaoqian M. and
                   Pajerowski, Daniel and Cheng, Yongqiang and Turner, Joshua J.
                   and Snyder, G. Jeffrey and Li, Mingda},
  title         = {Probing Non-Equilibrium Grain Boundary Dynamics with
                   XPCS and Domain-Adaptive Machine Learning},
  year          = {2026},
  eprint        = {2605.12194},
  archivePrefix = {arXiv},
  primaryClass  = {cond-mat.mtrl-sci},
  doi           = {10.48550/arXiv.2605.12194}
}

@article{Sutton2003,
  author  = {Sutton, Mark and Laaziri, Khalid and Livet, Fr{\'e}d{\'e}ric
             and Bley, Fran{\c c}ois},
  title   = {Using coherence to measure two-time correlation functions},
  journal = {Optics Express},
  volume  = {11},
  number  = {19},
  pages   = {2268--2277},
  year    = {2003},
  doi     = {10.1364/OE.11.002268}
}

@article{Madsen2010,
  author  = {Madsen, Anders and Leheny, Robert L. and Guo, Hongyu
             and Sprung, Michael and Czakkel, Orsolya},
  title   = {Beyond simple exponential correlation functions and equilibrium
             dynamics in X-ray photon correlation spectroscopy},
  journal = {New Journal of Physics},
  volume  = {12},
  number  = {5},
  pages   = {055001},
  year    = {2010},
  doi     = {10.1088/1367-2630/12/5/055001}
}

@article{RobinsonHarder2009,
  author  = {Robinson, Ian and Harder, Ross},
  title   = {Coherent X-ray diffraction imaging of strain at the nanoscale},
  journal = {Nature Materials},
  volume  = {8},
  pages   = {291--298},
  year    = {2009},
  doi     = {10.1038/nmat2400}
}

@article{Dargasz2026,
  author  = {Dargasz, Michelle and Anthuparambil, Nimmi Das and Retzbach, Sebastian
             and Girelli, Anita and Timmermann, Sonja and M{\"o}ller, Johannes
             and Jo, Wonhyuk and Leonau, Aliaksandr and Raza Agha, Mohammad
             and Bin, Maddalena and Savelkouls, Jaqueline and Andronis, Iason
             and Unger, Frederik and Brausse, Felix and Hallmann, J{\"o}rg
             and Boesenberg, Ulrike and Pudell, Jan-Etienne
             and Rodriguez-Fernandez, Angel and Wrigley, James
             and Shayduk, Roman and Youssef, Mohamed and Zozulya, Alexey
             and Madsen, Anders and Lehmk{\"u}hler, Felix and Perakis, Fivos
             and Zhang, Fajun and Schreiber, Frank and Paulus, Michael
             and Gutt, Christian},
  title   = {Depletion-induced interactions modulate nanoscale protein diffusion
             in polymeric crowder solutions},
  journal = {Proceedings of the National Academy of Sciences},
  volume  = {123},
  number  = {31},
  pages   = {e2524733123},
  year    = {2026},
  doi     = {10.1073/pnas.2524733123}
}

@article{Carr2021,
  author  = {Carr, Adra and Bowlan, John and Mazzoli, Claudio and Barbour, Andi
             and Hu, Wen and Wilkins, Stuart and Walker, Colby and Ding, Xiaxin
             and Kim, Jong Hyuk and Lee, Nara and Choi, Young Jai and
             Lin, Shi-Zeng and Sandberg, Richard L. and Zapf, Vivien S.},
  title   = {Evidence for the Alternating Next-Nearest Neighbor model in the
             dynamic behavior of a frustrated antiferromagnet},
  journal = {Physical Review B},
  volume  = {103},
  pages   = {L060401},
  year    = {2021},
  doi     = {10.1103/PhysRevB.103.L060401}
}

@article{Sheyfer2024,
  author  = {Sheyfer, Dina and Zheng, Hao and Krogstad, Matthew and
             Thompson, Carol and You, Hoydoo and Eastman, Jeffrey A. and
             Liu, Yuzi and Wang, Bi-Xia and Ye, Zuo-Guang and
             Rosenkranz, Stephan and Phelan, Daniel and Dufresne, Eric M. and
             Stephenson, G. Brian and Cao, Yue},
  title   = {X-ray-induced piezoresponse during X-ray photon correlation
             spectroscopy of {PbMg$_{1/3}$Nb$_{2/3}$O$_3$}},
  journal = {Journal of Synchrotron Radiation},
  volume  = {31},
  pages   = {55--64},
  year    = {2024},
  doi     = {10.1107/S1600577523009116}
}

@article{Carnis2019,
  author  = {Carnis, J{\'e}r{\^o}me and Gao, Lu and Labat, St{\'e}phane and
             Kim, Young Yong and Hofmann, Jan P. and Leake, Steven J. and
             Sch{\"u}lli, Tobias U. and Hensen, Emiel J. M. and
             Thomas, Olivier and Richard, Marie-Ingrid},
  title   = {Towards a quantitative determination of strain in Bragg coherent
             X-ray diffraction imaging: artefacts and sign convention in
             reconstructions},
  journal = {Scientific Reports},
  volume  = {9},
  pages   = {17357},
  year    = {2019},
  doi     = {10.1038/s41598-019-53774-2}
}

@article{Clark2015,
  author  = {Clark, Jesse N. and Ihli, Johannes and Schenk, Anna S. and
             Kim, Yi-Yeoun and Kulak, Alexander N. and Campbell, James M. and
             Nisbet, Gareth and Meldrum, Fiona C. and Robinson, Ian K.},
  title   = {Three-dimensional imaging of dislocation propagation during
             crystal growth and dissolution},
  journal = {Nature Materials},
  volume  = {14},
  pages   = {780--784},
  year    = {2015},
  doi     = {10.1038/nmat4320}
}

@article{Orr2023,
  author  = {Orr, Kieran W. P. and Diao, Jiecheng and
             Lintangpradipto, Muhammad Naufal and Batey, Darren J. and
             Iqbal, Affan N. and Kahmann, Simon and Frohna, Kyle and
             Dubajic, Milos and Zelewski, Szymon J. and Dearle, Alice E. and
             Selby, Thomas A. and Li, Peng and Doherty, Tiarnan A. S. and
             Hofmann, Stephan and Bakr, Osman M. and Robinson, Ian K. and
             Stranks, Samuel D.},
  title   = {Imaging Light-Induced Migration of Dislocations in Halide
             Perovskites with 3D Nanoscale Strain Mapping},
  journal = {Advanced Materials},
  volume  = {35},
  number  = {46},
  pages   = {2305549},
  year    = {2023},
  doi     = {10.1002/adma.202305549}
}

@article{ShaliziCrutchfield2001,
  author  = {Shalizi, Cosma Rohilla and Crutchfield, James P.},
  title   = {Computational Mechanics: Pattern and Prediction, Structure and Simplicity},
  journal = {Journal of Statistical Physics},
  volume  = {104},
  pages   = {817--879},
  year    = {2001},
  doi     = {10.1023/A:1010388907793}
}

@inproceedings{LittmanSuttonSingh2001,
  author    = {Littman, Michael L. and Sutton, Richard S. and Singh, Satinder},
  title     = {Predictive Representations of State},
  booktitle = {Advances in Neural Information Processing Systems},
  volume    = {14},
  pages     = {1555--1561},
  year      = {2001}
}

@inproceedings{TishbyPereiraBialek1999,
  author    = {Tishby, Naftali and Pereira, Fernando C. and Bialek, William},
  title     = {The Information Bottleneck Method},
  booktitle = {Proceedings of the 37th Annual Allerton Conference on
               Communication, Control, and Computing},
  pages     = {368--377},
  year      = {1999}
}

@article{BialekNemenmanTishby2001,
  author  = {Bialek, William and Nemenman, Ilya and Tishby, Naftali},
  title   = {Predictability, Complexity, and Learning},
  journal = {Neural Computation},
  volume  = {13},
  number  = {11},
  pages   = {2409--2463},
  year    = {2001},
  doi     = {10.1162/089976601753195969}
}

@article{Duri2005,
  author  = {Duri, Agn{\`e}s and Bissig, Hugo and Trappe, V{\'e}ronique
             and Cipelletti, Luca},
  title   = {Time-resolved-correlation measurements of temporally
             heterogeneous dynamics},
  journal = {Physical Review E},
  volume  = {72},
  pages   = {051401},
  year    = {2005},
  doi     = {10.1103/PhysRevE.72.051401}
}

@article{Wochner2009,
  author  = {Wochner, Peter and Gutt, Christian and Autenrieth, Tina
             and Demmer, Thomas and Bugaev, Volodymyr
             and Diaz Ortiz, Alejandro and Duri, Agnes and Zontone, Federico
             and Gruebel, Gerhard and Dosch, Helmut},
  title   = {X-ray cross correlation analysis uncovers hidden local symmetries
             in disordered matter},
  journal = {Proceedings of the National Academy of Sciences},
  volume  = {106},
  pages   = {11511--11514},
  year    = {2009},
  doi     = {10.1073/pnas.0905337106}
}

@article{Sanborn2011,
  author  = {Sanborn, Christopher and Ludwig, Karl F. and Rogers, Michael C.
             and Sutton, Mark},
  title   = {Direct Measurement of Microstructural Avalanches during the
             Martensitic Transition of Cobalt Using Coherent X-Ray Scattering},
  journal = {Physical Review Letters},
  volume  = {107},
  pages   = {015702},
  year    = {2011},
  doi     = {10.1103/PhysRevLett.107.015702}
}

@article{Konstantinova2021,
  author  = {Konstantinova, Tatiana and Wiegart, Lutz and Rakitin, Maksim
             and DeGennaro, Anthony M. and Barbour, Andi M.},
  title   = {Noise reduction in X-ray photon correlation spectroscopy with
             convolutional neural networks encoder--decoder models},
  journal = {Scientific Reports},
  volume  = {11},
  pages   = {14756},
  year    = {2021},
  doi     = {10.1038/s41598-021-93747-y}
}

@article{Konstantinova2022,
  author  = {Konstantinova, Tatiana and Wiegart, Lutz and Rakitin, Maksim
             and DeGennaro, Anthony M. and Barbour, Andi M.},
  title   = {Machine Learning for analysis of speckle dynamics:
             quantification and outlier detection},
  journal = {Physical Review Research},
  volume  = {4},
  pages   = {033228},
  year    = {2022},
  doi     = {10.1103/PhysRevResearch.4.033228}
}

@article{Timmermann2022,
  author  = {Timmermann, Sonja and Starostin, Vladimir and Girelli, Anita
             and Ragulskaya, Anastasia and Rahmann, Hendrik and Reiser, Mario
             and Begam, Nafisa and Randolph, Lisa and Sprung, Michael
             and Westermeier, Fabian and Zhang, Fajun and Schreiber, Frank
             and Gutt, Christian},
  title   = {Automated matching of two-time X-ray photon correlation maps
             from phase-separating proteins with Cahn--Hilliard-type
             simulations using auto-encoder networks},
  journal = {Journal of Applied Crystallography},
  volume  = {55},
  pages   = {751--757},
  year    = {2022},
  doi     = {10.1107/S1600576722004435}
}

@article{Horwath2024,
  author  = {Horwath, James P. and Lin, Xiao-Min and He, Hongrui
             and Zhang, Qingteng and Dufresne, Eric M. and Chu, Miaoqi
             and Sankaranarayanan, Subramanian K. R. S. and Chen, Wei
             and Narayanan, Suresh and Cherukara, Mathew J.},
  title   = {AI-NERD: Elucidation of relaxation dynamics beyond equilibrium
             through AI-informed X-ray photon correlation spectroscopy},
  journal = {Nature Communications},
  volume  = {15},
  pages   = {5945},
  year    = {2024},
  doi     = {10.1038/s41467-024-49381-z}
}

@article{Perakis2017,
  author  = {Perakis, Fivos and Amann-Winkel, Katrin and Lehmk{\"u}hler, Felix
             and Sprung, Michael and Mariedahl, Daniel and Sellberg, Jonas A.
             and Pathak, Harshad and Sp{\"a}h, Alexander and Cavalca, Filippo
             and Schlesinger, Daniel and Ricci, Alessandro and Jain, Avni
             and Massani, Barbara and Aubree, Flore and Benmore, Chris J.
             and Pettersson, Lars G. M. and Nilsson, Anders},
  title   = {Diffusive dynamics during the high-to-low density transition in amorphous ice},
  journal = {Proceedings of the National Academy of Sciences},
  volume  = {114},
  pages   = {8193--8198},
  year    = {2017},
  doi     = {10.1073/pnas.1705303114}
}
\end{document}